\documentclass[12pt,preprint]{aastex}

\shortauthors{Glikman, Gregg, Lacy et al.}
\shorttitle{FIRST-2MASS Red Quasars}

\begin{document}
\title {FIRST-2MASS Sources Below the APM Detection Threshold:\\ A Population of Highly Reddened Quasars}

\author{Eilat Glikman\altaffilmark{1}}
\email{eilatg@astro.columbia.edu}
\author{Michael D. Gregg\altaffilmark{2,3}}
\email{gregg@igpp.ucllnl.org}
\author{Mark Lacy\altaffilmark{4}}
\email{mlacy@ipac.caltech.edu}
\author{David J. Helfand\altaffilmark{1}}
\email{djh@astro.columbia.edu}
\author{Robert H. Becker\altaffilmark{2,3}}
\email{bob@igpp.ucllnl.org}
\and
\author{Richard L. White\altaffilmark{5}}
\email{rlw@stsci.edu}

\altaffiltext{1}{Columbia Univeristy Department of Astronomy, 550 West 120th Street, New York, NY 10027}
\altaffiltext{2}{Department of Physics, University of California at Davis, 1 Shields Avenue, Davis, CA 95616}
\altaffiltext{3}{Institute of Geophysics and Planetary Physics, Lawrence Livermore National Laboratory, Livermore, CA, 94551}
\altaffiltext{4}{SIRTF Science Center, 120 E. California Ave., Pasadena, CA, 91125}
\altaffiltext{5}{Space Telescope Science Institute, 3700 San Martin Dr., Baltimore, MD 21218}

\begin{abstract}
We have constructed a sample of bright near-infrared sources which are 
detected at radio wavelengths but undetected on the POSS I plates in
order to search for a population of dust-obscured quasars. Optical and
infrared spectroscopic followup of the sample has led to the discovery of
seventeen heavily reddened quasars ($B-K>6.5$), fourteen of which are reported 
here for the first time.  This has allowed us to define
a region in the $R-K, J-K$ color plane in which $50\%$ of the radio-selected 
objects are highly reddened quasars. We compare the surface density of this 
previously overlooked population to that of UVX- radio-selected quasars, 
finding that they make up $\sim 20\%$ of the total quasar population 
for $K\lesssim 15.5$.
\end{abstract}

\section {Introduction}

Optical surveys for quasars have often exploited the ultraviolet-excess (UVX) and power-law nature of the typical quasar spectrum.  Such searches have yielded tens of thousands of quasars with broad emission lines and blue optical colors ($U-B \lesssim -0.4$).  Although all quasars {\it may} be intrinsically blue in color, the presence of dust along the line-of-sight to a quasar's central engine will redden the observed spectrum.  Furthermore, quasars that are dimmed due to extinction will be selected against in magnitude-limited optical surveys, further biasing quasar number counts and affecting both the observed quasar luminosity function and its apparent evolution over cosmic time. 

\citet{sf92} argue that the presence of dust is responsible for the redder than average spectral energy distributions of low-ionization broad absorption line (BAL) QSOs and has led to an underestimate of the fraction of these objects in the general quasar population.  Additionally, examples of low-ionization BAL QSOs with unusually high Balmer decrements and red optical colors have supported a model in which quasars are buried in thick, dusty shrouds during an early phase of evolution \citep{egami96,becker00,lacy02}

\citet{web95} have suggested that a large population of quasars has gone undetected as a result of such biased searches. They utilized a sample of sources from the Parkes catalog brighter than 0.5 Jy at 2.7 GHz with flat radio spectra ($\alpha>-0.5$, $S_\nu\propto\nu^\alpha$).  The $B-K$ colors of their quasars range between 1 and 8.  This is a much broader color distribution than that of optically selected quasars whose optical to near-infrared colors are virtually all in the range $1<B-K<3.5$. \citet{web95} argue that if dust-obscured quasars are extinguished by several magnitudes in the optical ($A_B\approx5^m$), then a magnitude-limited optical survey would only detect intrinsically luminous red quasars.  Fainter quasars, extinguished by dust, will fall below a survey's magnitude limit and will not be detected.  If the same amount of reddening occurs in radio-quiet quasars as is found in the 0.5~Jy sources, then this extinction has caused $80\%$ of quasars to be missed by optical surveys. \citet{mal97} arrived at a similar conclusion for gravitationally lensed quasars after finding much redder optical to near-infrared colors in radio-selected lenses compared to optically-selected ones.  They atribute the red colors partly to dust in the lensing galaxies and partly to dust intrinsic to the quasars.  \citet{ke99} searched for the radio-quiet analog of the red quasar population using X-ray selection from ROSAT archival data.  Their results suggest that only $3\% - 20\%$ of quasars at a given intrinsic X-ray flux are obscured.  Other studies have shown equally discrepant results leading to no firm conclusion on the matter \citep{bdm95,sk97,benn98}

Spectulation has also arisen as to the source of the reddening.  \citet{web95} argue that the reddening in their quasar sample is due to dust obscuration.  However, the case has also been made for a red synchrotron component contributing to the near-infrared flux and causing the red colors in flat-spectrum radio-selected quasars \citep{sr97,whiting01}. Other constributions to the spread in $B-K$ colors may come from contamination by host galaxy starlight \citep{benn98,masci98} and/or an intrinsic spread in the optical continuum slope \citep{richards03}.  The emergence of 2MASS, a large-area near-infrared survey, provides the opportunity for a flux-limited $K$-band survey with the potential to at least partially recover this elusive quasar population.  

In this paper we present a new approach to finding this obscured population of quasars by searching for sources that are detectable both in the radio and near infrared but are optically faint.  We present an empirical method using optical to near-infrared colors to select out high-probablility quasar candidates.  We provide an estimate for the spatial density of this missing population and compare it to the previously detected quasar spatial density function. In Section 2 we discuss the initial sample selection criteria, while our optical spectroscopic followup observations are presented in \S 3.  We go on to refine our color selection criteria (\S 4) and compare the surface density of the objects on the sky to previous quasar surveys (\S 5), allowing us to assess quantitatively the missing component of the population.  In Section 6 we discuss biases and incompletenesses in our, and previous, quasar surveys.  We conclude (\S 7) with a discussion of future plans to improve and broaden our understanding of the red quasar population.

\section {The Selection Process}

We describe here our method to search for optically obscured quasars, incorporating information from radio, infrared, and optical catalogs.  We begin by defining an initial sample using three criteria:  (1) the object must be in the FIRST catalog;  (2) it must be in the 2MASS catalog, and (3) it must be fainter than the plate limit of the first generation Palomar Observatory Sky Survey (POSS-I).  Table 1 summarizes the characteristics of the three surveys.

The FIRST Survey \citep{bwh95} is a radio imaging project intended to serve as the 20-cm wavelength counterpart to the Palomar Observatory Sky Survey (POSS).  The July 2000 catalog covered 7988 deg$^2$ of sky and contained 722,354 primarily extragalactic sources with subarcsecond positions.  The survey is ideal for selecting optically obscured quasars.  Its 1 mJy flux density limit gives us the advantage of detecting QSO's down into the radio-quiet regime \citep{greg96}, allowing a search for optically obscured quasars over a broad range of radio luminosities.  The positional accuracy of the survey ($<1\arcsec$) allows us to reliably find optical counterparts to sources and/or to create a list of objects without optical counterparts.  The optically faint radio sources are excellent candidates for highly obscured AGN.

The Two Micron All Sky Survey \citep{2mass} began imaging the sky in 1997 using two 1.3 meter telescopes at Mt. Hopkins, Az, and CTIO, Chile.  The final release has covered $95\%$ of the sky in three near-infrared wavebands: $J (1.25 \mu$m$)$, $H (1.65 \mu$m$)$, and $K_s (2.17 \mu$m$)$.  With 4\arcsec\ resolution and an astrometric accuracy of $\sim 0\farcs 5$, a signal-to-noise of 10 at $J=15.8$, $H=15.1$, and $K_s=14.3$ mag has been achieved for point sources.  These limits correspond to a flux density of about 1 mJy in each band.  Sources with fainter magnitudes down to $K\sim16$ are detected, but with lower signal-to-noise and considerable incompleteness.  The Second Incremental Release, which we have used in this experiment, reports a sky coverage of 19,681 deg$^2$ containing more than 160 million point sources \footnote{See 2MASS home page at http://www.ipac.caltech.edu/2mass/index.html}.  

The 2MASS Second Incremental Release is made up of 27,493 tiles.  Each tile represents a $6\degr$ long scan from south to north and is $8.5\arcmin$ wide in right ascension.  The overlap between FIRST and 2MASS is broken up by these tiles which are not necessarily adjacent.  To determine the coverage area of our survey we counted the FIRST sources that fell inside the four corners of each 2MASS tile.  Since the FIRST sources are extragalactic and uniformly distributed on the sky, the fraction of FIRST sources that fall inside a 2MASS tile represents the fraction of the FIRST coverage area included in both surveys.  A total of 244,988 FIRST sources lie in the 2MASS survey area ($34\%$ of the total July 2000 FIRST catalog), establishing our survey size to be 2716 deg$^2$.

We matched the FIRST sources with the 2MASS catalog out to offsets of 12\arcsec\ in order to determine the optimal search radius for selecting counterparts.  We also matched the 2MASS catalog to a false FIRST catalog generated by shifting the entire FIRST catalog by 5\arcmin\ to the north.  Figure 1 shows a histogram of FIRST-2MASS separations.  Integrating under the two curves out to 2\arcsec\ shows that $93.6\%$ of our matches to this radius are physically associated.  This yields an optically unbiased sample of 12,435 sources (denoted F2M hereafter).  Given the relatively low resolution of 2MASS data, we restricted our matches to objects with a stellar 2MASS morphological classification (ext\_flg = 0). 

The Cambridge Automated Plate Machine \citep{apm92} has digitized the POSS-I photographic plates at a resolution of 0\farcs5.  The catalog produced by these scans includes all objects detected down to the plate limits, $O\sim21.5$ and $E\sim20$.  We used the APM-FIRST catalog\footnote{The APM-FIRST catalog is available at the FIRST home page \url{http://sundog.stsci.edu}} which contains all matched objects with separations up to 20\arcsec\ \citep{mc02} to search for unidentified objects from our F2M source list.  We matched our F2M list to the APM-FIRST catalog with a search radius of 2\arcsec.  A total of 10,500 sources had matches in the APM catalog, leaving 1,935 objects unmatched.  A false catalog, produced by shifting F2M by 5\arcmin\ to the north, produced 33 matches with a separation closer than 2\arcsec, a $0.3\%$ false rate.

Several factors cause optical sources associated with a F2M object to be mistakenly listed as 'unmatched.'  The most common error is caused by objects that lie in parts of the sky not included in the APM catalog.  We used the GSC2.2 catalog to help solve this problem for sources brighter than $F\sim18.5$ and $J\sim19.5$.

The Guide Star Catalog II (GSC II) is an all-sky optical catalog from the second generation Palomar Observatory and UK Schmidt Sky Surveys.  A subset of this effort, GSC2.2, is a magnitude-limited catalog which has been made publicly available.  It contains about 450 million objects brighter than 18.5 in photographic $F$ and 19.5 in photographic $J$.  We used a subset of this catalog, produced by matching FIRST against GSC 2.2 out to radius of 24\arcsec.  We matched this catalog, containing 292,032 sources, to the 1,935 F2M objects with no APM counterpart to remove bright optical sources.  We used a matching radius of 2\arcsec\ which yielded 1354 matches. A false catalog, produced by shifting F2M by 5\arcmin\ to the north produced 3 matches with a separation closer than 2\arcsec. Since our objective was to find optically {\it faint} sources, our interest lies with the remaining 581 F2M sources with neither APM nor GSC2.2 counterparts.

Sources fainter than the GSC2.2 magnitude limit and brighter than the POSS-I plate limit  had to be removed by hand.  We conducted a visual inspection of all 581 sources using postage stamp cutouts from the first generation digitized sky survey to select sources which were not visible or are very faint on the plate images.  We then compared the image statistics of both the undetected, APM- scanned sources and the unscanned, visually inspected sources.  Using a $10\times10$ pixel square ($17\arcsec\times17\arcsec$) around the center of each APM-scanned source, we used the IRAF task IMSTAT to measure the brightest pixel and the number of standard deviations ($t\sigma$) above background, to quantify the sensitivity of the APM scans.  Figure 2 shows the distribution of $t$ (with a gaussian fit overplotted) for the sources scanned but undetected by APM.  We repeated these measurements on the sources from regions of the sky not covered in the APM catalog and removed sources which were brighter than $4\sigma$ above background, which, as shown in Figure 3, implies a $95\%$ probability that the APM machine would not have detected them.  We also removed from our sample the six sources in Figure 3 that had counts more than $4\sigma$ above the background.  

 We also found, through visual inspection, that another common error involved F2M objects whose separations from their optical counterparts were slightly greater than 2\arcsec.  We removed by hand galaxies, several arcseconds in size, whose nominal optical positions were more than 2\arcsec\ away from the radio positions, but which were clearly associated with the same object.  For fainter sources (APM $R \gtrsim 18.5$), we used statistics derived from \citet{mc02} to estimate the probability that a nearby APM source was in fact associated with the F2M source.  The probabilities of real association ranged from $12\%$ to $95\%$ depending on the separation and APM morphology of the source.  We then summed the probabilities of association $P$ starting with the lowest until the total reached $383\%$; the addition of these ten sources to the sample would thus mean the mistaken inclusion of $\sim 3.8$ real optical detections to the list of putatively undetected radio-IR sources. The sum of $(1-P)$ for the remaining objects was $341\%$, implying an incompleteness of $\sim 3.4$ sources (objects that we take to be optically identified radio-IR sources but which are in fact chance coincidences; all of these had probabilities of matching of $>75\%$).  The final sample including the ten low-probability matches then includes 69 objects and is $\sim 95\%$ reliable ($1-3.8/69$) and $\sim 95\%$ complete ($1-3.4/69$).

\section{Optical and Near-Infrared Spectroscopic Observations}

Spectroscopy of our candidate quasars was carried out in the optical and near-infrared at four different observatories.  The majority of the optical spectra were obtained at the 10m Keck telecopes using the ESI and, to a lesser extent, LRIS spectrographs. Optical spectra were also obtained at the 3 m Shane telescope at Lick Observatory.  Near-infrared spectroscopy was performed at the NASA Infrared Telescope Facility (IRTF) using SpeX \citep{spex1} and at the MDM Observatory on Kitt Peak using the TIFKAM infrared camera in spectroscopic mode \footnote{TIFKAM was funded by the Ohio State University, the MDM consortium, MIT, and NSF grant AST-9605012. NOAO and USNO paid for the development of the ALADDIN arrays and contributed the array currently in use in TIFKAM.}.  We also obtained seven spectra from the Sloan Digital Sky Survey Spectral database. 

The LRIS data were reduced and spectra extracted using standard IRAF procedures.  ESI spectra were produced using a combination of IRAF tasks and customized software developed by the authors.  More details on the algorithms used in the ESI software are given in \citet{white03a}.

The near infrared spectra obtained at IRTF with SpeX were reduced using Spextool \citep{spex3}.  We flux calibrated and corrected the data for telluric absorption using spectra of nearby A0V stars observed immediately before or after each object spectrum, following the procedure outlined in \citet{spex2}.  The near infrared spectra obtained with TIFKAM at MDM were extracted and reduced using standard spectral reduction tasks in the IRAF package ONEDSPEC.  We used both A0V and G2V stars (to calibrate the hydrogen absorption lines in the A0V star) when correcting these spectra for telluric absorption, following \citet{hanson96}.

We classified our spectra using algortihms in the radial velocity analysis (RV) package of IRAF.  We used the FXCOR task to cross-correlate the spectra with a set of galaxy and star templates, implementing the technique illustrated in \citet{td79}.  Our templates consisted of elliptical, S0, Sa, Sb, Sc and starburst galaxies from \citet{kin96} and \citet{cal94}\footnote{We chose the starburst galaxy template with $0.61 < E(B-V) < 0.70$.  Of the six available templates, this spectrum had the highest reddening;  it was chosen as most consistent with our survey's selection criteria.}.  This task was most useful in classifying and assigning redshifts for the absorption line galaxies.  We used the Tonry \& Davis $R$ value to determine the strength of the correlation, rejecting identifications with $R<3.0$.  Some spectra with emission features were misidentified as starbursts, but were assigned correct redshifts; others did not correlate at all.  In the latter cases we used the RV task RVIDLINES to measure radial velocities from spectral lines using emission features present in the quasar composite spectrum from \citet{broth01} as well as the first three Paschen lines.  We then fit gaussian profiles to the emission lines and classifed objects with line widths $> 1000$ km/s as quasars.  Objects with narrow lines were classified as starbursts or narrow-line AGN based on [O III]/H$\beta$ and [N II]/H$\alpha$ line diagnostics \citep{oster89}.

Table 2 summarizes all the data collected on our quasar candidates.  Columns (1) and (2) list the candidates' J2000 coordinates from the FIRST catalog.  In columns (3) and (4) we list the $B$ and $R$ magnitudes, respectively, from the complete Guide Star 2 catalog which reaches the POSS-II plate limits.  We chose to present these magnitudes despite our selection criteria which required that the objects be fainter than the APM limit, in order to allow for near-infrared to optical color studies which we discuss in the following sections.  In columns (5), (6) and (7) we present 2MASS $J$,$H$, and $K_s$ photometry, followed by the peak and integrated FIRST 20 cm flux densities in columns (8) and (9).  In columns (10) and (11) we present $J-K$ and $R-K$ colors.  We used the $R$ magnitude from GSC2 for $R-K$.  We argue in \S 4 that these near-infrared and optical to near-infrared colors can be useful diagnostics for efficient selection of red quasars.  We note, however, that the non-simultaneity of the POSS-II and 2MASS observations make our colors provisional at best.  We report the redshift of the candidate when available in column (12), followed by the classification we determined using the aforementioned analysis in column (13); any other references in the literature are provided in column (14).  

\section{Color Space Selection Criteria for Red Quasars}

We have obtained optical and/or near-infrared spectra for 50 of the 69 optically faint F2M candidates.  Twenty-three of these objects are some kind of AGN, eighteen of which are bona fide quasars (with line widths $\gtrsim 1000$ km$s^{-1}$).  One of these quasars has a blue optical spectrum and may have been included due to variability between the POSS-I and 2MASS epochs or photometric errors in one of the surveys (SDSS photometry is similar to the GSC2 values).  Nineteen are galaxies (eighteen red ellipticals and one emission line galaxy), and eight are stars, mostly type M or later. The spectra for the quasars are displayed in Figure 3.  We scaled the optical spectrum to the near infrared spectrum by measuring the median value in their overlapping spectral region ($\lambda = 8900-10000$\AA) and scaling the spectrum with the lower median flux to that with the higher median flux.  This approach to combining the optical and near-infrared spectra is correct in the absence of spectral variability between the epochs of optical and near-infrared observations; it also assumes that there are no significant color gradients in the sources which would produce color changes in the face of differing amounts of slit losses during the inevitable changes of observing conditions from run to run.

In Figure 4 we plot the 69 sources in $J-K$ vs. $R-K$ color space.  As expected, there are more red quasars ($\circ$) at redder $J-K$ and $R-K$ colors.  More interestingly, galaxies ($+$) tend to be found at bluer $R-K$ and slightly bluer $J-K$ colors.  Stars ($\ast$), on the other hand, have rather red $R-K$ colors similar to those of the red quasars, but are {\it much} bluer in $J-K$.  The AGN ($\triangle$) occupy the same part of this space as the quasars.  

Two of the red quasars, F2M J082502.0+471651 and F2M J165647.1+382136 (plotted in Figure 3), showed only narrow emission lines in their Keck ESI spectra, but had clear broad Paschen-$\beta$ in the near-infrared.  Had we not observed these objects in the near infrared, they would have been classified as narrow-line AGN.  Another narrow-line AGN for which we obtained both an optical and a near-infrared spectrum, F2M J073806.0+214103, showed only narrow Pa-$\alpha$ in the $K$-band.  However, it is quite possible that one or more of the four remaining narrow-line AGN found in our survey which have no corresponding near-infrared spectrum may have broad Paschen lines and thus may in the future be promoted to red quasars (cf. Rawlings et al.\ 1995).

Of the nineteen galaxies identified in this survey, only three have both an optical and a near-infrared spectrum: elliptical galaxies F2M J011712.7-012755 and F2M J144812.1+305614, and the emission line galaxy F2M J141736.2+225346.  The remaining sixteen objects only have optical spectra.  Since it is possible for a quasar to be reddened sufficiently that its host galaxy light dominates the optical spectrum, the lack of a near-infrared spectrum which could reveal broad Paschen lines can lead to an incorrect classification (similar to the case of the narrow-line AGN discussed above).  Furthermore, galaxy light in such an object can affect its optical to near-infrared colors, preventing it from being targeted as a quasar candidate in a color-color selection scheme (see below).

If we separate the populations with a color cut at $J-K=1.7$ and $R-K=4$ (the dotted lines in Figure 4) we can analyze our quasar detection efficiency in color-color space.  There are 31 objects with $J-K>1.7$ and $R-K>4$.  Thirty of these objects have been classified: 14 red quasars ($46.7\%$), 1 blue quasar ($3.3\%$), 5 narrow-line AGN ($16.7\%$), 9 galaxies ($30\%$), and 1 star ($3.3\%$).  With these color criteria we have a success rate of nearly $50\%$ for finding red quasars.  

For the 38 objects with $J-K<1.7$ or $R-K<4$ we have obtained 20 identifications: 1 emission-line galaxy ($5\%$), 9 galaxies ($45\%$), 7 stars ($35\%$), and 3 red quasars.  All three quasars are consistent within the errors of lying inside the boundaries of the red quasar color space defined above.  It is clear, therefore, that adopting our color cuts produces a high efficiency for detecting red quasars and largely avoids galaxies and stars.

The $J-K$ color region in which the quasars in this survey lie is consistent with the results of \citet{cutri01} who serched for red AGN using a $J-K>2$ color cut on objects in the 2MASS Point Source Catalog with no optical or radio constraints and found $\sim 150$ red AGN, mostly with $z \leq 0.5$.  Only two of these quasars have $z>1.5$; F2M J092145.6+191812 is recovered in this survey.  \citet{bh01} use quasars from the \citet{vcv00} quasar catalog with 2MASS counterparts to argue that a two-color selection method (with one of the colors extending over a long wavelength baseline) is an efficient way to select quasars. The addition of a second color-constraint may explain the higher redshift objects found in this survey. 

Our results are also consistent with the ideas put forth by \citet{whf00} who argue that although optically red stars (such as the M stars detected in our survey) and red quasars will have similar spectral energy distributions in the optical, the blackbody shape of the stellar spectrum will deviate from the power law shape of the quasar spectrum in the infrared.  They suggest using optical to infrared colors (such as our $R-K$) in conjunction with infrared colors (such as $J-K$) to separate stars from quasars in a manner insensitive to reddening. 

We note that the fraction of spectroscopically identified objects in our sample depends rather strongly on $K$ magnitude. Figure 5 shows a histogram of the $K$ magnitudes binned by 0.25 magnitudes.  The unshaded histogram includes all 69 candidates.  We overplot a histogram shaded with vertical lines for the 50 spectroscopically identified objects.  The bins containing the 17 red quasars are colored black.  We have identifications for $80\%$ of the objects brighter than $K=15.5$, but only $38\%$ for $K>15.5$.  Although the efficiency of finding red quasars decreases for objects with smaller $R-K$ values (which necessarily have fainter $K$ magnitudes due to our $R>21$ restriction), it is possible that there are more quasars at fainter $K$ magnitudes.  Therefore, we restrict our comparison between this sample and optically selected samples to objects brighter than $K=15.5$ and account for this incompleteness in the following discussion.

\section{The Observed Surface Density of Quasars on the Sky}

We wish to compare the spatial density of the red quasars in our sample to the general quasar population in order to determine the fraction of quasars missed by UVX selection methods.  Using a Small Magellanic Cloud (SMC) dust reddening law from \citet{gc98}, we find that a $z=0.5$ quasar with $E(B-V)=0.5$ will suffer $\sim 2$ magnitudes of extinction in the observed-frame $B$-band while in the observed-frame $K$-band the amount of extinction is only $\sim 0.2$ magnitudes.  At $z=2.5$ the extinction in $K$ rises to $\sim 1.3$ magnitudes compared with $\sim 9$ magnitudes of extinction in $B$.  Therefore, if we compare spatial density distributions of quasars as a function of $K$ magnitude, as opposed to the traditional $B$ magnitude, we largely avoid comparing luminous dust-extinguished (and therefore optically faint) quasars with unextinguished blue quasars with much lower luminosities.

Although the surface density of UVX-selected quasars has been determined as a function of $B$ magnitude or its equivalent \citep{hs90,lfc97} no such determination has been published to date in the $K$ band. In order to construct an optically selected sample that is comparable to our red quasars, we need to compare our sample with quasars whose optical magnitude limit approaches the plate limit of POSS-I and which are also detected in FIRST and 2MASS. Figure 6 compares the sky density of our F2M red quasars to quasars selected in optical magnitude-limited surveys.  We describe below the method by which we determine the $K$ magnitude surface density on the sky for quasars.

The First Bright Quasar Survey (FBQS) selects quasars by matching FIRST sources to the APM catalog.  The resultant samples are magnitude-limited in the optical photographic $E$ (red) band and obey the color criterion $O-E<2$.  We calculated their surface density as a function of $K$ magnitude by matching the FBQS to the 2MASS All Sky Data Release. 

The FBQS II \citep{fbqs2} contains 636 quasars in 2682 deg$^2$ at an optical limit of $E = 17.8$ magnitudes in the APM catalog. There were 503 2MASS matches to FBQS II.  Its optically fainter analog, FBQS III \citep{ffqs}, contains 321 quasars in 589 deg$^2$ and has a magnitude limit of $E = 18.9$. There were 134 2MASS detections in this sample.  In Figure 6 we plot the surface density of these samples separately, as a function of $K$ magnitude, in 0.5 magnitude bins.  FBQS II and III are represented by $+$ and $\times$ symbols, respectively. 

We also constructed a UVX-selected sample by combining the Large Bright Quasar Survey (LBQS) described in \citet{lbqs95} and the 2dF QSO Redshift Survey (2QZ) described in \citet{cr01}. We use the subset of FIRST- and 2MASS-detected sources within these quasar catalogs to serve as a UVX-selected sample that is comparable to the F2M red quasars. The LBQS survey contains 58 such detections in 270 deg$^2$ in the optical magnitude range $16.0 \leq B_J \lesssim 18.85$.  The 2QZ sample was matched to FIRST by \citet{cir03} who found 104 matches in 122 deg$^2$.  This list had nine matches in 2MASS in the optical range $18.25 \leq B_J \leq 20.85$.  Removing the two of these already listed in LBQS, yields a total sample of 65 unique UVX-selected quasars with FIRST and 2MASS matches.  We combine this sample and scale the number counts by their respective coverage areas in order to have a continuous UVX-selected sample that extends effectively to the plate limit of POSS-I ($\diamond$ symbols in Figure 6).

Since the final release of 2MASS covers the whole sky, the unmatched quasars from the above samples must have $K$ magnitudes fainter than the 2MASS detection limit, and hence would not have been detected in any quasar survey based on 2MASS detections such as our optically faint F2M sample.  Therefore, comparing the spatial density of these known quasars in the $K$ band using a FIRST plus 2MASS filter enables us to compare samples with uniform characteristics.

By comparing the distribution functions in Figure 6 we can determine the fraction of quasars that a $K$ magnitude-limited sample would recover versus one that is optically selected.  We see that radio-selected and UVX-selected, radio-detected quasars have comparable surface densities in the $K$ band.  The shapes of the optically bright and optically faint (F2M) quasar distributions are also similar, but the F2M blank field sample constitutes only a small fraction of the blue quasars.

As a further check that we are comparing objects from the same overall population, we plot in Figure 7 the redshift distribution of the F2M blank field survey quasars along with the FBQS II and III quasars that have been detected by 2MASS.  All three surveys span the same redshift range, and none shows a correlation of reddening with redshift.   The slight upturn at redder $B-K$ colors toward low redshifts in the FBQS II sample is likely due to added $K$-band light from the quasars' host galaxies.  

We list the QSO surface densities from the optical surveys along with those of the F2M sample in Table 3.  We derive the uncertainties using Poisson statistics for the FBQS II and III as well as the UVX selected quasars.  Since the F2M sample has identifications for only $72\%$ of the candidates ($80\%$ for $K\leq 15.5$) we must adjust the error bars on our spatial density measurements. We account for the incomplete spectroscopic identifications by making the most conservative assumption possible: that all the unobserved objects are quasars.  We make this the upper limit on top of the Poisson uncertainties for the quasar counts in this sample.  The last line in the Table gives the integrated surface density of all quasars brighter than $K=15.5$.  If the red quasars in the F2M blank field sample are red due to dust extinction, then the optical surveys are missing objects whose extinction-corrected magnitudes would put them above the survey threshold, leading to an underestimate of the population.  If we assume that these are otherwise normal quasars, then the $K$-band surface density in each optical survey ($n_{opt}$) {\it plus} the surface density of the F2M blank field red quasars ($n_{\mathrm{F2M}}$) ought to represent the total surface density of quasars.  By adding up the surface densities from the Table we can determine what fraction of this total is made up of red quasars missed by optically selected samples; the percentage of red quasars is

\begin{equation}
\frac{n(K\leq15.5)_{\mathrm{F2M}}}{n(K\leq15.5)_{opt} + n(K\leq15.5)_{\mathrm{F2M}}} \times 100 .
\end{equation}

Applying this to the data in Table 3, we determine that the UVX selection method misses $2.5\pm0.9\%$ of the quasars in a $K$-magnitude-limited sample.  The FBQS II and III miss $3.7\pm1.2\%$ and $3.9\pm1.4\%$, respectively.

These are strict lower limits to the fractions of missing red quasars, since both the UVX method and the FBQS surveys impose a color criterion on their candidates as well, and it is possible that there are red quasars detected on the POSS-I plates redder than the FBQS color cut of $O-E=2$.  These would be luminous optical objects, which may experience several magnitudes of extinction, yet still be bright enough to show up on POSS-I.  The F2M blank field sample thus does not contain all the red quasars in the FIRST-2MASS list and we are currently working on a program to create a more complete list of red quasars.  We note, however, that the percentages of missed quasars are unlikely to change by more than a factor of four or so once this intermediate population is identified (see below).  The F2M blank field sample reported here is unique in that it contains both very highly reddened objects as well as quasars that come from the fainter end of the optical luminosity function and have been reddened just enough to make it into our survey.  

\section{Biases in Quasar Surveys Due to Reddening}

We have compared the surface densities of optically selected, radio-detected quasars to near-infrared-selected, radio-detected quasars as a function of apparent $K$ magnitude.  The fraction of missing quasars derived from this comparison is valid for such magnitude-limited surveys.  Eventually, however, we would like to be able to make statements about the fraction of red quasars in the overall quasar population, and how the missing group of objects identified here affects the luminosity function of quasars.  In this case it is not enough to use apparent magnitudes. In the following section we discuss the limitations of magnitude-limited surveys in the near-infrared and the optical wavelength regimes due to redshift and reddening.  

\subsection{Surveys in the Near-Infrared}

Though the effects of dust are reduced when number counts on the sky are compared in the infrared, the effect of dust on the detectability of red quasars in the near-infrared increases with redshift.  If the dust is at the same redshift as the quasar then rest frame optical light from high redshift ($z\gtrsim 1.3$) quasars will be shifted into the $H$ and $K$-bands and will generate incompleteness even in a near-infrared survey.  We will find highly reddened low-redshift quasars, but as we look at higher redshifts we will be biased toward lower extinctions and intrinsically more luminous objects.  This bias was noted by \citet{white03} in an $I$-band survey for quasars which found five quasars with $E(B-V)>0.2$, all at $z<1.3$.  These were also the most luminous quasars at their respective redshifts, after dereddening.

To determine the reddening of the quasars in the F2M sample, we fit a quasar composite spectrum which spans UV to near-infrared rest wavelengths to our red quasar spectra. We use the FBQS composite spectrum \citep{broth01} for the UV and optical wavelengths  $800-8000$\AA.  We also obtained near-infrared spectra at IRTF with SpeX for ten ordinary quasars chosen from the Sloan Digital Sky Survey Quasar Catalog for which an absorption of zero has been assumed.  These quasars have $K$-magnitudes ranging between 12 and 14.5 and are in the redshift range $0.16<z<0.38$.  We Doppler corrected the spectra, normalized them to the mean flux in the $1.3-1.4 \mu$m wavelength range and averaged them (Glikman et al., in preparation).   We scaled the near-infrared composite to the optical composite by matching the flux in the wavelength range $7000-8000$\AA\ and attaching them at $7500$\AA, to create one spectrum extending from $800-20800$\AA.  We followed the method outlined in \citet{white03} to redden this composite with a Small Magellanic Cloud dust law from \citet{gc98} and fit it to our spectra, thereby determining $E(B-V)$.  We then deredden the seventeen F2M red quasars and plot their intrinsic absolute $K$-band magnitudes as a function of redshift in Figure 8.  We compare them to the absolute magnitudes of FBQS II and III quasars for which no intrinsic absorption has been assumed.  

The five red quasars found by \citet{white03} were the most luminous quasars in their sample.  Furthermore they are all found at relatively low redshift as a result of the detection limit of the survey.  Figure 8 shows that in the case of near-infrared-selected quasars we are able to detect highly reddened objects $E(B-V)=0.3-1.0$ with typical absolute $K$-band luminosities out to $z\sim1.3$.  At higher redshifts however, near-infrared selection also begins to be affected by extinction.  The red quasars detected at $z>1.3$ are the most luminous (as well as having the least amount of extinction).  In a survey with no optical constraints, with a given $K$-band magnitude limit, our data suggest that we would be able to detect most or all of the red quasars at low redshifts, but at higher redshifts our detection limit would prevent us from finding highly reddened quasars with 'normal' intrinsic luminosities.  The dashed lines in the Figure show our detection limits for quasars at a given absolute $K$ magnitude, redshift, and extinction.  Redder quasars must be more luminous intrinsically to make it into a magnitude-limited survey, especially at $z>1.5$.  To recover these objects a deeper $K$-band survey is necessary.

The trend of finding red quasars with smaller extinctions and higher intrinsic luminosities with increasing redshift is clear from Figure 8.  However the two highest redshift, most luminous quasars (F2M J013435.6-093102 and F2M J100424.8+122922) are both gravitationally lensed, and hence their high absolute magnitudes are partly due to magnification and should be treated as upper bounds.  Furthermore, in the case of F2M J013435.6-093102, \citet{hall02} present a reddening model which places the dust in a $z=0.76$ lensing galaxy and results in $E(B-V) = 1.3$.  Our composite fit places the dust at the quasar's redshift and computes $E(B-V) = 0.47$ (lower than the value of 0.67 found by \citet{hall02}). 

In Table 4 we list the parameters extracted from our composite fits to the seventeen F2M quasars which we order by decreasing redshift.  Columns (1) and (2) list the quasars' J2000 coordinates.  Column (3) lists the 2MASS apparent $K$ magnitude followed by $A_K$, the amount of absorption in $K$, in column (4).  Column (5) lists the fitted $E(B-V)$ used in shading the points in Figure 8.  In column (6) we present the radio-loudness parameter followed by the redshift in column (7).  

For a given total extinction, the amount of absorption in the observed $K$-band increases with redshift.  Correcting the $K$ magnitudes of the F2M quasars by their fitted $A_K$ values shifts the apparent $K$ magnitudes by $\sim$0.3-1.2 magnitudes and lowers the intrinsic magnitude limit of our survey to $\lesssim$15.0 (ignoring the nineteen candidates for which we do not have spectra).  If we apply the spatial densities of the seventeen F2M quasars (0.006 deg$^{-2}$) and quasars from FBQS I and II brighter than $K=15.0$ (0.103 deg$^{-2}$ and 0.08 deg$^{-2}$, respectively) to equation (1) we find that the percentage of red quasars in this intrinsic-magnitude-limited sample increases.  The FBQS II and III therefore miss $\gtrsim 5.5-7\%$. 

The radio-loudness distribution of the F2M quasars listed in column (7) of Table 4 was calculated using the definition  $R\equiv f(1.4\mathrm{GHz})/f(B_{intrinsic})$\footnote{\citep{sto92} define $R$ as the ratio of radio to optical power:  $R=f(5$GHz$)/f(B)$. The ``radio-loud'' regime is entered when $R>2.0$, while ``radio-quiet'' objects are generally designated as those with $R<0.5$.}  without including a $K$-correction.  We computed $B_{intrinsic}$ by correcting the $B$ magnitude from Table 2 by the $A_B$ values which we obtained from the composite fits.  An $R$ value of 10 divides the radio-loud and radio-quiet regimes, though objects in the range $3 \lesssim R \lesssim 100$ are seen as radio intermediate.  The majority of FBQS sources lie in this range, as do about two thirds of the F2M quasars.  

In this work, we have assumed that the red quasars in the F2M sample are dust reddened. However, the spectral slope in the radio may provide further information on the source of a red quasar's optical to near-infrared colors.  \citet{sr97} argued that enhanced synchrotron emission from relativistic jets lying in the line of sight contributes to a quasar's near-infrared flux and is responsible for the large spread of $B-K$ colors in the Parkes half-Jansky flat-radio-spectrum quasar sample of \citet{web95}.  \citet{whiting01} looked for a synchrotron signature in the spectral energy distributions of the Parkes red quasars and found evidence for a red synchrotron component in $\sim 40\%$ of the sources.  To determine if the quasars in our survey have flat radio spectra, we are obtaining contemporaneous 3.6 cm and 20 cm flux density measurements of the 17 quasars in our survey with the VLA, we reserve a detailed study of the reddening for a future publication (Glikman et al., in preparation).

\subsection{Optical Surveys}

Unextinguished, normal blue quasars have a typical $B-K \sim 2.5$ as was demonstrated by \citet{web95} using the colors of 26 LBQS quasars.  We find this to be equally true for the FBQS II and III samples.  In Figure 9 we plot the $K$ band surface density function for the FBQS II and III quasars listed in Table 3.  These points are plotted with $\diamond$ and $\square$ symbols, repectively, and connected with lines.   We also calculated and plotted ``fake'' $K$ band magnitudes from the $O$ magnitudes listed in the FBQS II and III catalogs ($K_{fake} = O - 2.5$, $\ast$ and $\times$  symbols, respectively).  Between $13 \leq K \leq 15.5$ the real and fake points are remarkably close.  The discrepancies at $K<13.5$ arise due to the small number of bright quasars.  At fainter magnitudes the functions diverge due to incompleteness in the 2MASS catalog.  The fake points therefore have slightly higher spatial densities at fainter $K$ since they were calculated from their measured optical magnitudes.  

This further bolsters the case that optical selection of quasars produces a uniform subgroup of objects not representative of the overall quasar population: it misses red quasars.  In Figure 10 we plot the $B-K$ color distribution of the FBQS II and III quasars that were detected by 2MASS.  We scaled the number counts to each survey's respective coverage area so that we are again comparing surface densities, this time as a function of color.  We compare this to the color distribution of the F2M red quasars which seems to be the tail end of the $B-K$ color distribution for FIRST-2MASS quasars.  The dearth of objects at $5 \lesssim B-K \lesssim 6.5$ is due to our survey's own selection criterion which requires that objects be too faint to show up on the POSS plates.  This restriction automatically forces $B-K \ga 5$.  

At a given extinction in the observed $B$-band, ($A_B$), we can determine the intrinsic $B$ magnitude range of quasars that would fall within and outside the F2M blank field survey's optical constraints.  If we assume that the intrinsic color of most quasar continua is $B-K\simeq 2.5$ then the intrinsic optical magnitude limit of the F2M blank field survey, given the 2MASS limit of $K\lesssim 15.5$, is $B\lesssim 18$.  Because these objects are not detected on the POSS-I plates, $A_B\geq 3$.  It is possible, however, for a $B=15.5$ magnitude object that is extinguished by 3 magnitudes to be missed by this survey, since it {\it will} appear on the POSS-I plates.  UVX selection methods will likely not find it because of its colors.  Figure 10 shows that FBQS III does detect redder quasars than FBQS II due to its deeper optical limit.  However, an FBQS equivalent survey extending to the APM detection limit is unlikely recover any new 2MASS detected sources.  These objects would necessarily have $B-K\geq 5.4$, conflicting with the survey's $O-E<2$ color constraint.  

Any incompleteness in the red quasar population that arises due to our optical constraints can be recovered in a survey for red quasars using the color selection discussed in \S 4, without regard to their detection or non-detection by the APM scans of POSS-I.  These intermediate objects should fill in the gap between the FBQS and F2M $B-K$ color distributions.  

We have begun a program to search for objects detected on the POSS-I plates using the $R-K > 4$ and $J-K>1.7$ color selection criteria developed here.  We use preliminary results from this program to estimate the number of quasars that would be added to our survey and have found 137 objects in FIRST-2MASS, including the thirty-one objects found in this survey, meeting the aforementioned color criteria.  We use the GSC2.3 catalog to obtain the $R$ magnitudes (photographic $F$) for the $R-K$ color cuts.    Since two of our red quasars fall outside this region, this expanded catalog contains fifteen of the seventeen red quasars listed here.  If we assume that half of these objects will be quasars, as was true for the optically faint subset, then we expect to find a total of about seventy red quasars in the same region of sky. This is roughly a factor of four more quasars than were found in this survey, which raises the percentage of red quasars in the overall quasar population to between $~10\%$ and $15\%$ in a magnitude-limited sample.  As we showed in Section 6.1, these percentages should increase by a factor of about 1.5 after correcting for absorption in the $K$-band.

\section{Summary and Future Work}

We have conducted a survey to search for optically obscured, infrared-bright quasars, detected in both the FIRST and 2MASS surveys.  We required that the objects be undetected by the APM scans of the POSS-I plates.  Using these criteria we obtained a list of 69 candidates.  Of the 51 spectroscopically observed candidates, 17 were red quasars. From the optical and near infrared colors of the candidates  we have shown that red quasar detection efficiency rises to $\sim 50\%$ when adopting the color criteria $R-K>4$ and $J-K>1.7$.  

While the near-infrared is helpful in finding red quasars because of its weaker sensitivity to dust extinction, its effectiveness diminishes with redshift and increased obscuration.  To evaluate the significance of red quasars to the whole quasar population it is necessary to characterize the distribution of $E(B-V)$ as a function of intrinsic luminosity and redshift.  With a sample of seventeen quasars this cannot yet be accomplished.   We have made some estimates of the percentage of red quasars which range from a firm lower limit from directly detected and classified optically faint objects of $\sim 3\%$, to an extinction- and survey-bias-corrected best estimate of $\sim 20\%$.

Future work will include a wider survey for red quasars taking advantage of the all-sky coverage of the latest 2MASS release.  We will also eliminate optical magnitude constraints and will select candidates based solely on their optical and near-infrared colors.   We intend to study the reddening of these objects in greater detail by calculating Balmer decrements and comparing unreddened spectra to quasar composites made from optically selected samples to search for any intrinsic differences between the two populations.  

We thank Dave Deutsch of the Manhattan Center for Science and Mathematics for his careful and consistent inspection of the Digitized Sky Survey images to weed out optical identifications of red quasar candidates.  E. G. and D. J. H. acknowledge support from National Science Foundation grant AST-00-98259.  R. H. B. acknowledges support of the Institute of Geophysics and Planetary Physics (operated under the auspices of the US Department of Energy by the University of California, Lawrence Livermore National under contract W-7405-Eng-48) as well as additional support from NSF grant AST00-98355 (University of California, Davis).

This publication makes use of data products from the Two Micron All Sky Survey, which is a joint project of the University of Massachusetts and the Infrared
Processing and Analysis Center/California Institute of Technology, funded by the National Aeronautics and Space Administration and the National Science
Foundation.
 
The Guide Star Catalogue-II is a joint project of the Space Telescope Science Institute and the Osservatorio Astronomico di Torino. Space Telescope Science Institute is operated by the Association of Universities for Research in Astronomy, for the National Aeronautics and Space Administration under contract NAS5-26555. The participation of the Osservatorio Astronomico di Torino is supported by the Italian Council for Research in Astronomy. Additional support is provided by European Southern Observatory, Space Telescope European Coordinating Facility, the International GEMINI project and the European Space Agency Astrophysics Division. 

Funding for the Sloan Digital Sky Survey (SDSS) has been provided by the Alfred P. Sloan Foundation, the Participating Institutions, the National Aeronautics and Space Administration, the National Science Foundation, the U.S. Department of Energy, the Japanese Monbukagakusho, and the Max Planck Society. The SDSS Web site is \url{http://www.sdss.org/}.

The SDSS is managed by the Astrophysical Research Consortium (ARC) for the Participating Institutions. The Participating Institutions are The University of Chicago, Fermilab, the Institute for Advanced Study, the Japan Participation Group, The Johns Hopkins University, Los Alamos National Laboratory, the Max-Planck-Institute for Astronomy (MPIA), the Max-Planck-Institute for Astrophysics (MPA), New Mexico State University, University of Pittsburgh, Princeton University, the United States Naval Observatory, and the University of Washington.

\begin{deluxetable}{cccccc}
\tablewidth{0pt}
\tablehead{
\colhead{Survey} & \colhead{Wavelength} & \colhead{Sensitivity} & \colhead{Resolution} & \colhead{Number of Sources} & \colhead{Coverage Area\tablenotemark{a}} }
\tablecaption{Summary of Catalog Characteristics \label{catsum}}
\startdata
FIRST & 21 cm & 1 mJy & 1\arcsec & 722,354 & 7,988 deg$^2$ \\ 
2MASS & J(1.25\micron)    & 15.8 mag\tablenotemark{b}& 4\arcsec & 162,213,354 & 19,600 deg$^2$ \tablenotemark{c} \\
 & H(1.65\micron) & 15.1 mag\tablenotemark{b} & & & \\
 & K$_s$(2.17\micron) & 14.3 mag\tablenotemark{b} & & & \\
GSC2.2 & photographic F(6750\AA) & 18.5 mag & 1\arcsec & 435,457,355 & All Sky \\
 & photographic J(5750\AA) & 19.5 mag & & & \\
APM & photographic E(6500\AA) & $\sim$20 mag & 1\arcsec & 514,815\tablenotemark{d} & 15,000 deg$^2$ \\
 & photographic O(4000\AA) & $\sim$21.5 mag & & & \\
\enddata
\tablenotetext{a}{For catalog versions used in this work; Subsequent work has expanded FIRST and 2MASS coverage.}
\tablenotetext{b}{For S/N = 10}
\tablenotetext{c}{The 2MASS data release used here has a patchy sky coverage due to the survey's scanning strategy.  See the 2MASS home page for details \url{http://www.ipac.caltech.edu/2mass/}}
\tablenotetext{d}{Catalog from \citet{mc02} APM-FIRST match to 20\arcsec.}
\end{deluxetable}

\begin{deluxetable}{ccrrcccrrccccl}
\tabletypesize{\scriptsize}
\rotate
\tablewidth{0pt}
\tablehead{
\colhead{RA} & \colhead{Dec} & 
\colhead{$B$\tablenotemark{a}} & \colhead{$R$\tablenotemark{a}} & 
\colhead{$J$} & \colhead{$H$} & \colhead{$K$} & \colhead{$F_{pk}$} & \colhead{$F_{int}$} 
&\colhead{$J-K$} & \colhead{$R-K$} &\colhead{$z$} & \colhead{Type} & \colhead{Ref.}\\
\colhead{(2000)} & \colhead{(2000)} & \colhead{(mag)} & \colhead{(mag)} & 
\colhead{(mag)} & \colhead{(mag)} & \colhead{(mag)} & \colhead{(mJy)} &
\colhead{(mJy)} & \colhead{(mag)} & \colhead{(mag)} &\colhead{} & \colhead{} & 
\colhead{}\\
\colhead{(1)} & \colhead{(2)} & \colhead{(3)} & \colhead{(4)} & 
\colhead{(5)} & \colhead{(6)} & \colhead{(7)} & \colhead{(8)} & \colhead{(9)} 
&\colhead{(10)} & \colhead{(11)} &\colhead{(12)} & \colhead{(13)} & \colhead{(14)}
}
\tablecaption{Optically Faint FIRST-2MASS Red Quasar Candidates}
\startdata
00 00 25.51&$-$09 57 53.0&   22.11&   19.20&16.94&16.42&15.64& 3.64& 9.18&1.30&\phs3.56&0.354&Galaxy&\tablenotemark{c}\\
00 43 29.18&$-$10 10 36.9&   21.48&   18.64&17.07&15.89&15.34& 9.62&15.28&1.73&\phs3.30&0.479&Galaxy&\tablenotemark{c}\\
00 44 02.81&$-$10 54 18.9&   22.34&   19.10&17.77&16.10&15.06&38.60&42.57&2.71&\phs4.04&0.431&Galaxy&\\ 
01 00 23.33&$-$10 08 01.3&$>$22.50&   20.09&16.62&16.11&15.91& 1.22& 1.03&0.71& $>$4.18&     &M-star&\tablenotemark{c}\\
01 02 01.42&$-$08 40 14.6&   22.33&   19.53&16.88&16.41&15.71& 1.13& 1.31&1.17&\phs3.82& & & \\
01 17 12.76&$-$01 27 55.4&$>$22.50&$>$20.80&17.28&16.36&15.33& 8.34& 8.66&1.94&\phs5.47&0.864&Galaxy&\\
01 25 15.98&$-$09 16 58.0&   21.53&   18.63&16.64&16.17&15.25& 1.40& 1.77&1.39&\phs3.38&0.389&Galaxy&\tablenotemark{c}\\
01 28 37.01\tablenotemark{d}&$+$00 02 38.9\tablenotemark{d}&   17.78&   16.33&16.70&16.08&15.17& 5.33&10.16&1.53&\phs1.16&0.392&Galaxy&\tablenotemark{c}\\
01 33 19.43&$-$00 56 06.6&   20.27&   18.35&16.81&16.09&15.43& 2.90& 4.25&1.38&\phs2.92& & & \\ 
01 34 35.68&$-$09 31 03.0&$>$22.50&$>$20.80&16.17&14.75&13.55&900.38&920.52&2.62&$>$7.25&2.220&QSO&\citet{greg02} \tablenotemark{c}\\
01 59 58.28&$-$05 04 27.7&   21.61&   19.18&16.83&16.22&15.55& 1.55& 1.05&1.28&\phs3.63& & & \\
02 01 05.69&$-$03 17 56.3&   21.70&   19.16&17.04&16.25&15.68& 1.24& 1.68&1.36&\phs3.48& & & \\
02 02 59.77&$-$09 23 44.7&   21.65&   19.17&16.74&15.74&15.27& 2.86& 2.45&1.47&\phs3.90&0.333&Galaxy&\tablenotemark{c} \\
02 08 04.41&$-$03 23 49.5&   21.51&   18.60&17.03&16.07&15.28& 4.13& 5.08&1.75&\phs3.32&0.442&Galaxy& \\
02 09 54.35&$-$04 13 41.0&   20.23&   18.31&16.69&15.77&15.52& 1.16& 2.09&1.17&\phs2.79& & & \\
02 14 50.32&$-$03 44 58.0&   21.52&   18.90&16.88&16.48&15.46& 1.04& 0.74&1.42&\phs3.44& & & \\
02 45 34.43&$-$07 00 53.3&   21.30&   18.67&16.87&16.24&15.47& 1.64& 1.09&1.40&\phs3.20& & & \\
07 05 25.70&$+$43 59 36.7&   20.71&   18.63&17.23&15.98&15.84&15.41&15.65&1.39&\phs2.79& & & \\
07 24 23.98&$+$26 26 44.4&   21.27&$>$20.80&17.27&16.10&15.22& 1.71& 1.90&2.05& $>$5.58&0.308&Galaxy& \\
07 28 12.39&$+$22 40 37.9&$>$22.50&   19.39&18.63&16.25&15.45&19.33&19.93&3.18&\phs3.94&0.272&QSO& \\
07 33 39.64&$+$45 25 21.5&$>$22.50&   19.09&16.86&16.06&15.04& 1.69& 2.28&1.82&\phs4.05&0.453&Galaxy& \\
07 38 06.06&$+$21 41 03.1&$>$22.50&$>$20.80&17.13&16.23&15.03& 2.28& 1.76&2.10& $>$5.77&0.275&NL AGN& \\
07 38 20.10&$+$27 50 45.5&$>$22.50&$>$20.80&17.05&16.17&15.26& 2.64& 2.33&1.80&\phs5.54&1.985&QSO&\citet{greg02}\\
08 12 29.11&$+$25 07 31.0&$>$22.50&   19.17&17.18&16.12&15.50& 1.38& 0.66&1.68&\phs3.67& & & \\
08 25 02.05&$+$47 16 52.0&   22.36&   20.45&17.01&15.79&14.11&61.12&63.24&2.90&\phs6.34&0.803&QSO&\tablenotemark{b} \\
08 34 07.01&$+$35 06 01.8&   21.51&   18.96&16.55&15.72&14.65& 1.22& 0.55&1.89&\phs4.31&0.470&QSO&\\
08 39 54.87&$+$17 12 48.3&$>$22.50&$>$20.80&16.58&15.94&16.09& 2.50& 2.46&0.49& $>$4.71& & & M-star\\
08 41 04.98&$+$36 04 50.1&$>$22.50&$>$20.80&17.53&16.31&14.92& 6.49& 6.58&2.62& $>$5.88&0.553&QSO&\\
08 55 55.69&$+$36 47 51.5&$>$22.50&$>$20.80&16.93&16.14&15.25& 1.16& 1.34&1.68& $>$5.55& & &\\
09 00 22.82&$+$20 11 11.6&$>$22.50&$>$20.80&16.83&16.36&15.75& 1.30& 1.50&1.08& $>$5.05& & M star \\
09 06 51.52&$+$49 52 36.0&   20.82&$>$20.80&17.06&15.83&15.14&67.87&75.11&1.91& $>$5.66&1.635&QSO&\\
09 08 49.78&$+$13 25 15.9&$>$22.50&   19.05&16.64&15.91&15.35& 1.33& 2.03&1.29&\phs3.70& & & \\
09 21 45.69&$+$19 18 12.6&   21.32&   20.12&16.75&15.43&14.55& 4.40& 4.72&2.20&\phs5.57&1.791\tablenotemark{e}&QSO&\citet{smith02}\\
09 27 15.23&$+$16 07 58.9&$>$22.50&$>$20.80&16.46&15.64&15.80& 4.42&11.24&0.66& $>$5.00& & & \\
09 34 15.06&$+$46 55 04.5&   21.14&   19.77&16.47&15.93&15.27& 5.06&11.31&1.20&\phs4.50& & M-star & \\ 
09 44 51.93&$+$31 13 52.8&$>$22.50&   19.65&17.01&16.23&15.28& 1.52& 1.33&1.72&\phs4.37&0.512&Galaxy& \\
09 46 42.55&$+$18 40 34.0&$>$22.50&$>$20.80&19.03&17.95&15.46& 3.69& 5.00&3.57& $>$5.34&0.403&NL AGN&\\
09 49 09.62&$+$18 56 47.9&   21.28&   18.57&17.34&16.15&15.43& 2.54& 3.11&1.91&\phs3.14&0.434&Galaxy& \\
09 50 32.44&$+$18 52 22.8&$>$22.50&   19.64&17.34&16.15&15.43& 1.51& 1.40&1.90&\phs4.21&0.480&Galaxy&\\
09 54 38.79&$+$17 35 24.8&   21.46&   19.43&16.67&15.86&15.77& 1.55& 1.07&0.90&\phs3.66& & & \\
09 56 23.12&$-$00 10 22.2&   21.28&   19.07&17.25&16.14&15.40& 1.86& 3.17&1.85&\phs3.67& & & \\
09 58 53.63&$+$12 38 23.0&   20.74&   19.34&16.76&15.80&14.74& 3.31& 3.25&2.02&\phs4.60& & & \\
10 04 24.87&$+$12 29 22.4&$>$22.50&$>$20.80&16.55&15.50&14.54&11.42&12.32&2.01& $>$6.26&2.650&QSO&\citet{lacy02}\\
10 12 30.48&$+$28 25 27.2&$>$22.50&$>$20.80&17.33&16.92&15.27& 9.23& 8.76&2.06& $>$5.53&0.937&QSO& \\
10 15 28.64&$+$12 07 51.9&   22.15&   19.56&17.32&16.48&15.30&10.60&10.98&2.02&\phs4.26&0.426&NL AGN& \\
10 22 29.39&$+$19 29 38.9&   20.94&   19.20&16.91&16.08&15.12& 2.09& 2.41&1.78&\phs4.08&0.406&NL AGN& \\
10 49 18.24&$+$15 44 58.1&$>$22.50&$>$20.80&17.86&16.76&15.36& 5.33& 5.42&2.49& $>$5.44&0.677&Galaxy& \\
11 18 11.06&$-$00 33 41.9&$>$22.50&   19.87&17.04&15.64&14.58& 1.30& 1.81&2.46&\phs5.29&0.686&QSO& \\
11 51 24.07&$+$53 59 57.4&   21.92&$>$20.80&17.11&16.28&15.10& 3.52& 3.60&2.00& $>$5.70&0.780&QSO& \\
11 57 33.67&$+$16 11 39.5&$>$22.50&$>$20.80&16.33&15.73&15.29& 2.19& 2.97&1.04& $>$5.51& &M-star& \\
12 02 55.33&$+$26 15 18.8&   20.69&   19.34&17.27&16.42&15.19&43.01&44.25&2.08&\phs4.15&0.565&NL AGN& \\
12 08 27.75&$+$17 08 14.1&   22.11&   19.40&16.62&15.26&15.32& 2.82& 2.86&1.30&\phs4.08&0.549&Galaxy& \\
12 19 03.10&$+$19 05 12.8&$>$22.50&   19.35&16.93&15.62&15.29& 5.10& 4.99&1.64&\phs4.06&1.37&QSO& \\
12 23 02.79&$+$17 52 46.8&   22.12&   19.63&16.71&16.10&15.52& 1.57& 5.80&1.19&\phs4.11& &M star& \\
12 29 24.80&$+$25 09 21.0&$>$22.50&   20.21&16.78&16.34&15.43& 2.07& 1.80&1.35&\phs4.78& & & \\
13 00 53.51&$+$19 01 02.5&   22.03&   18.69&17.12&15.99&15.93& 2.12& 1.79&1.19&\phs2.76& & & \\
13 05 26.13&$-$03 54 40.2&   20.95&   18.76&17.49&15.92&15.38& 9.66&14.40&2.11&\phs3.38&0.424&Galaxy& \\
13 41 08.11&$+$33 01 10.3&$>$22.50&$>$20.80&16.93&15.55&14.85&69.41&69.81&2.08& $>$5.95&1.720&QSO& \\
13 53 08.65&$+$36 57 51.2&$>$22.50&$>$20.80&17.41&15.58&14.26& 3.71& 3.32&3.15& $>$6.54&1.311&QSO& \\
13 59 41.18&$+$31 57 40.5&   21.63&   19.56&16.93&15.84&14.79& 1.33& 1.28&2.14&\phs4.77&1.724&QSO& \\
14 09 08.09&$+$52 11 32.1&   21.14&   18.68&15.83&15.24&14.75& 8.15& 8.60&1.08&\phs3.93& & & \\
14 17 36.23&$+$22 53 46.9&$>$22.50&$>$20.80&16.38&15.58&14.77& 1.66& 1.11&1.61& $>$6.03& 0.362&H {\tiny II} Galaxy& \\
14 22 46.15&$+$24 04 11.4&$>$22.50&$>$20.80&17.53&16.17&15.33& 1.87& 1.35&2.20& $>$5.47& &M-star& \\
14 48 12.15&$+$30 56 14.5&   22.02&   19.39&16.67&16.20&14.96& 1.02& 0.94&1.71&\phs4.43&0.271&Galaxy& \\
15 07 18.10&$+$31 29 42.3&   21.69&   19.87&16.79&15.95&15.17& 7.79& 7.73&1.62&\phs4.70&0.99&QSO& \\
15 57 18.75&$+$38 08 50.9&$>$22.50&   19.20&16.69&16.24&15.37& 1.10& 0.71&1.32&\phs3.83& & & \\
16 56 47.11&$+$38 21 36.7&$>$22.50&$>$20.80&17.30&17.72&15.09& 4.12& 3.89&2.21& $>$5.71&0.732&QSO&\tablenotemark{b}\\
22 24 38.35&$-$00 07 50.9&$>$22.50&   19.78&16.81&15.70&15.02& 1.41& 1.29&1.79&\phs4.76&0.452&Galaxy&\\
23 38 32.94&$-$10 22 38.8&   22.41&   19.83&16.51&15.78&15.27& 2.71& 2.56&1.24&\phs4.56& &star& \\
\enddata
\tablenotetext{a}{Photographic $F$ and $J$ magnitudes determined by GSC2 (the full version, without the magnitude limits mentioned in Table 1)}
\tablenotetext{b}{This object was initially classified as an emisson line galaxy, based on its optical spectrum.  The near-infrared spectrum reveals a red continuum and a broad $Pa-\beta$ line. See \S 4 for further discussion.}
\tablenotetext{c}{SDSS optical spectrum.}.
\tablenotetext{d}{This object had a FIRST-APM separation of 5.77\arcsec\ and was identified as an extended source by the APM machine's scans of both the $O$ and $E$ plates.  These data gave the objects a $54\%$ probability of being associated with each other.  Our criterion of $94\%$ reliability, discussed in \S 2, allowed this source to remain a candidate.  Its extended radio morphology as well as optical magnitudes and spectrum suggest that this object is the contaminant allowed by our survey statistics.}
\tablenotetext{e}{We determined the redshift from H$\alpha$ which lies between the $H$ and $K$ bands; its signal was strong enough to be detected in the region of telluric absorption.  We also used a weaker H$\beta$ line also residing in the telluric absorption region between $J$ and $H$.  This allowed us to identify Mg II absorption in the optical spectrum at this same redshift. \citet{smith02} obtained an independent redshift measurement of $z=1.800$ for this object using H-$\alpha$ alone.}
\end{deluxetable}

\begin{deluxetable}{ccccc}
\tablecolumns{5}
\tablewidth{0pt}
\tablecaption{QSO Counts and Surface Densities in $K$-Band}
\tablehead{ 
\colhead{Magnitude Range} & \colhead{FBQS II} & 
\colhead{FBQS III} & \colhead{UVX\tablenotemark{a}} & 
\colhead{F2M} \\
\colhead{$K$} & \multicolumn{4}{c}{(deg$^{-2}$ 0.5 mag$^{-1}$)} 
}
\startdata
11.0 - 11.5 & 0.0011$\pm$0.0006 & \nodata         & \nodata        & \nodata \\
11.5 - 12.0 & 0.0004$\pm$0.0004 & 0.002$\pm$0.002 & \nodata        & \nodata \\
12.0 - 12.5 & 0.003$\pm$0.001   & \nodata         & \nodata        & \nodata \\
12.5 - 13.0 & 0.004$\pm$0.001   & \nodata         & \nodata        & \nodata \\
13.0 - 13.5 & 0.006$\pm$0.002   & \nodata         & \nodata        & \nodata \\
13.5 - 14.0 & 0.014$\pm$0.002   & 0.007$\pm$0.003 & 0.007$\pm$0.005& 0.0004$\pm$0.0004 \\
14.0 - 14.5 & 0.030$\pm$0.003   & 0.031$\pm$0.007 & 0.04$\pm$0.01  & 0.0007$^{+0.0007}_{-0.0005}$ \\
14.5 - 15.0 & 0.044$\pm$0.004   & 0.042$\pm$0.009 & 0.07$\pm$0.02  & 0.003$^{+0.004}_{-0.001}$ \\
15.0 - 15.5 & 0.052$\pm$0.004   & 0.07$\pm$0.01   & 0.12$\pm$0.04  & 0.003$^{+0.009}_{-0.001}$\\
15.5 - 16.0 & 0.030$\pm$0.003   & 0.07$\pm$0.01   & 0.04$\pm$0.02  &$<0.009$ \\
16.0 - 16.5 & 0.002$\pm$0.001   & 0.010$\pm$0.004 & \nodata        &\nodata \\
\hline
\hline
                  & (deg$^{-2}$)    & (deg$^{-2}$)  & (deg$^{-2}$)  & (deg$^{-2}$)   \\
\phm{0.00} - 15.5 & 0.154$\pm$0.008 & 0.15$\pm$0.02 & 0.23$\pm$0.05 & 0.006$\pm$0.002\\
\enddata
\tablenotetext{a}{This sample is made up of the LBQS and 2QZ samples combined as described in \S 5.1}
\end{deluxetable}

\begin{deluxetable}{cccccrc}
\tablecolumns{7}
\tablewidth{0pt}
\tablecaption{Extinction Parameters from Composite Fits to F2M Red Quasars}
\tablehead{ 
\colhead{RA} & \colhead{Dec} & \colhead{$K$} & \colhead{$A_K$} &
\colhead{$E(B-V)$} & \colhead{R} & \colhead{$z$} \\
\colhead{(2000)} & \colhead{(2000)} & \colhead{(mag)} & \colhead{(mag)} & 
\colhead{(mag)} & \colhead{} & \colhead{}\\
\colhead{(1)} & \colhead{(2)} & \colhead{(3)} & \colhead{(4)} & 
\colhead{(5)} & \colhead{(6)} & \colhead{(7)}
}
\startdata
10 04 24.87 & $+$12 29 22.4 & 14.54 & 1.09 & 0.40 &   $>$3.2 & 2.650 \\
01 34 35.68 & $-$09 31 03.0 & 13.55 & 1.08 & 0.47 & $>$295.4 & 2.220 \\
07 38 20.10 & $+$27 50 45.5 & 15.26 & 0.91 & 0.45 &   $>$2.3 & 1.985 \\
09 21 45.69 & $+$19 18 12.6 & 14.55 & 1.02 & 0.56 &      0.5 & 1.791 \\
13 59 41.18 & $+$31 57 40.5 & 14.79 & 0.87 & 0.49 &      0.6 & 1.724 \\
13 41 08.11 & $+$33 01 10.3 & 14.85 & 0.86 & 0.49 &  $>$66.9 & 1.720 \\
12 19 03.10 & $+$19 05 12.8 & 15.29 & 0.52 & 0.38 &  $>$35.3 & 1.370 \\
13 53 08.65 & $+$36 57 51.2 & 14.26 & 1.18 & 0.89 &   $>$0.2 & 1.311 \\
15 07 18.10 & $+$31 29 42.3 & 15.17 & 0.51 & 0.50 &     19.4 & 0.990 \\
10 12 30.48 & $+$28 25 27.2 & 15.27 & 0.62 & 0.64 &  $>$19.1 & 0.937 \\
08 25 02.05 & $+$47 16 52.0 & 14.11 & 0.69 & 0.80 &     53.2 & 0.803 \\
11 51 24.07 & $+$53 59 57.4 & 15.10 & 0.54 & 0.64 &      7.0 & 0.780 \\
16 56 47.11 & $+$38 21 36.7 & 15.09 & 0.63 & 0.78 &   $>$6.1 & 0.732 \\
11 18 11.06 & $-$00 33 41.9 & 14.58 & 0.47 & 0.61 &   $>$6.9 & 0.686 \\
08 41 04.98 & $+$36 04 50.1 & 14.92 & 0.64 & 0.95 &   $>$7.0 & 0.553 \\
08 34 07.01 & $+$35 06 01.8 & 14.65 & 0.31 & 0.51 &      8.3 & 0.470 \\
07 28 12.39 & $+$22 40 37.9 & 15.45 & 0.41 & 0.84 & $>$104.3 & 0.272 \\
\enddata
\end{deluxetable}

\begin{figure}
\figurenum{1}
\epsscale{1}
\plotone{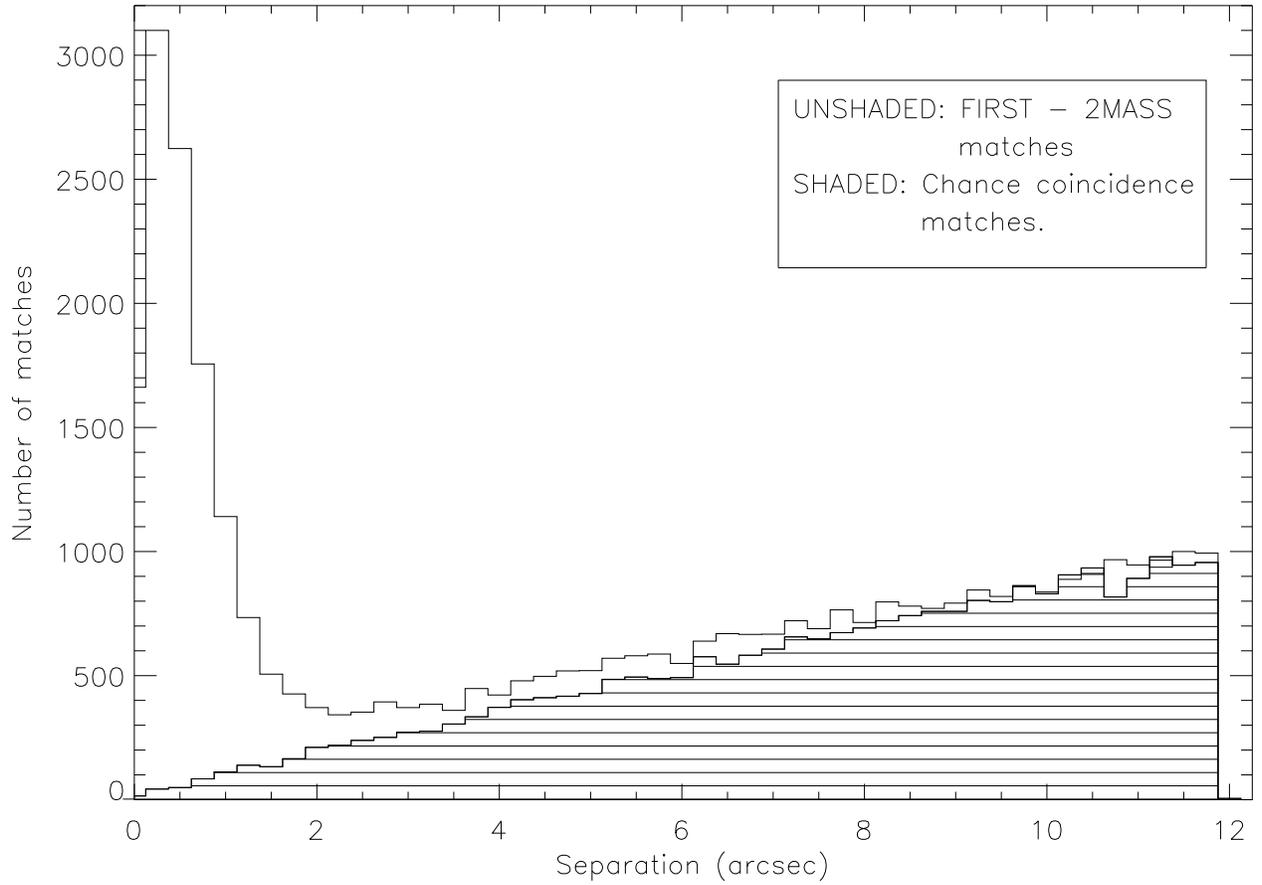}
\caption{Histogram of separations for F2M matches, binned into 0\farcs25 bins. Only the closest 2MASS source to each FIRST source is included.  The shaded histogram shows the number of matches to the false FIRST catalog in each bin.}
\end{figure}

\begin{figure}
\figurenum{2}
\plotone{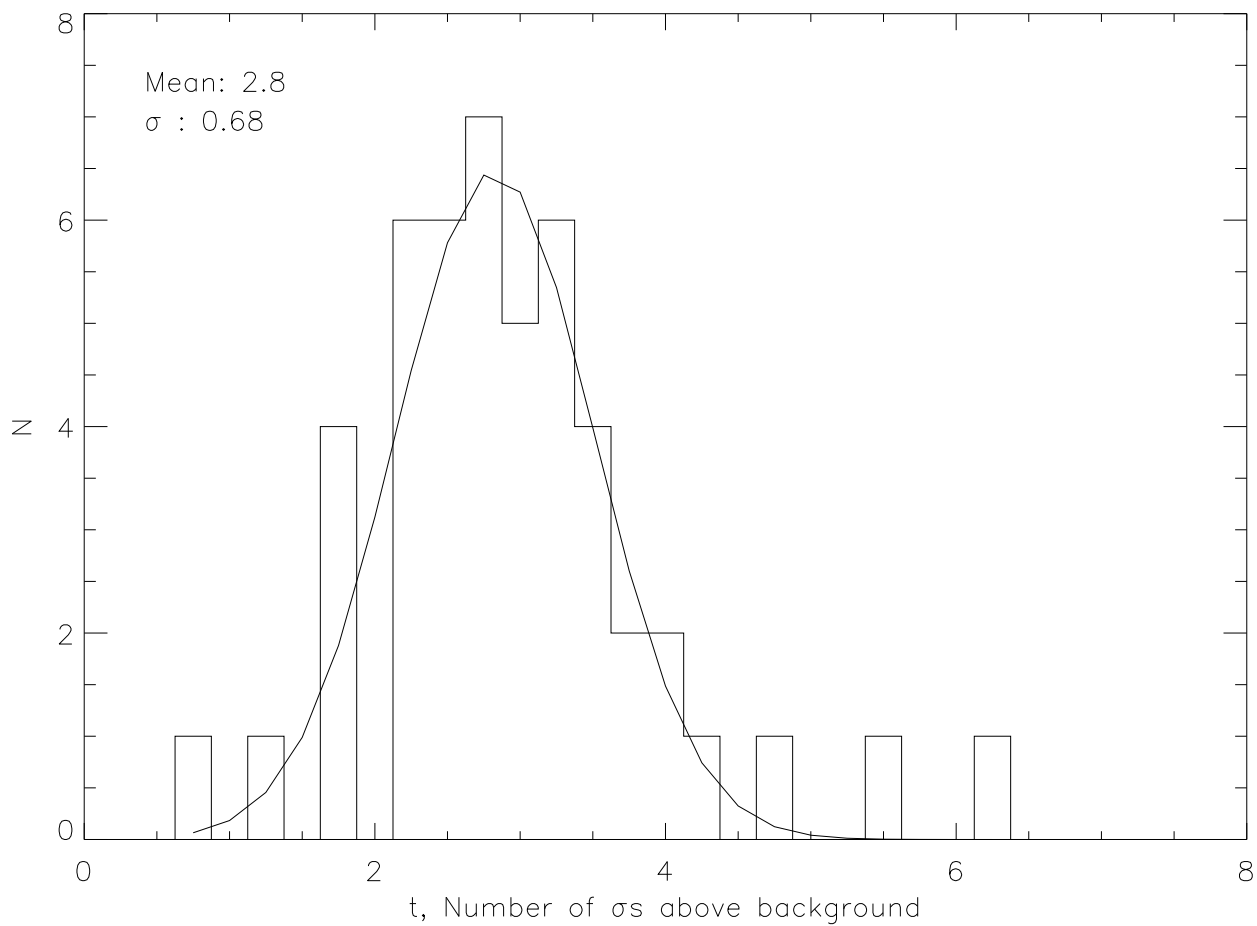}
\caption{Sources undetected by the APM plate scans.  The Digitized Sky Survey provides information about the strength of a source, in terms of standard deviations from the background, $t\sigma$.  This plot shows a histogram of $t$ values for the ``undetected'' F2M sources.  Clearly, many of these objects are present on the plates just below the APM detection threshold.  Sources that are fainter than $4\sigma$ above the background have a $95\%$ probability (i.e., two standard deviations from the mean of {\it this} distribution) of being undetected by the APM machine. The six sources brighter than this limit have been dropped from our candidate list. }
\end{figure}

\begin{figure}
\figurenum{3}
\plotone{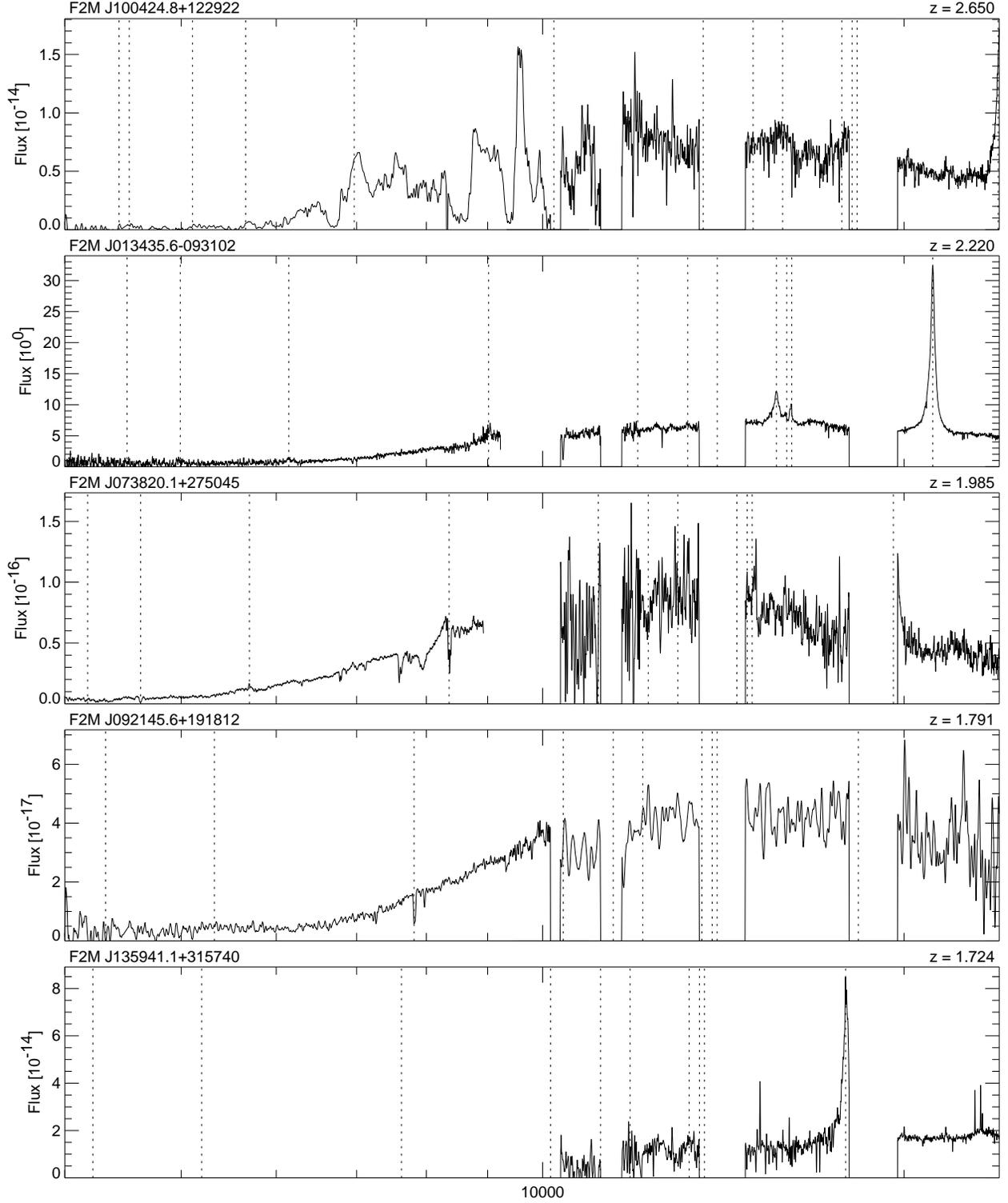}
\caption{
Spectra of eighteen F2M candidates identified as quasars, ordered by
redshift.  The dotted lines show expected positions of
prominent emission lines in the optical and near-infrared:
Ly$\alpha$~1216,
N~V~1240,
Si~IV~1400,
C~IV~1550,
C~III]~1909,
Mg~II~2800,
[O~II]~3727,
H$\delta$~4102,
H$\gamma$~4341,
H$\beta$~4862,
[O~III]~4959,
[O~III]~5007,
H$\alpha$~6563,
Pa$\gamma$~10941,
Pa$\beta$~12822.
}
\end{figure}

\begin{figure}
\figurenum{3b}
\plotone{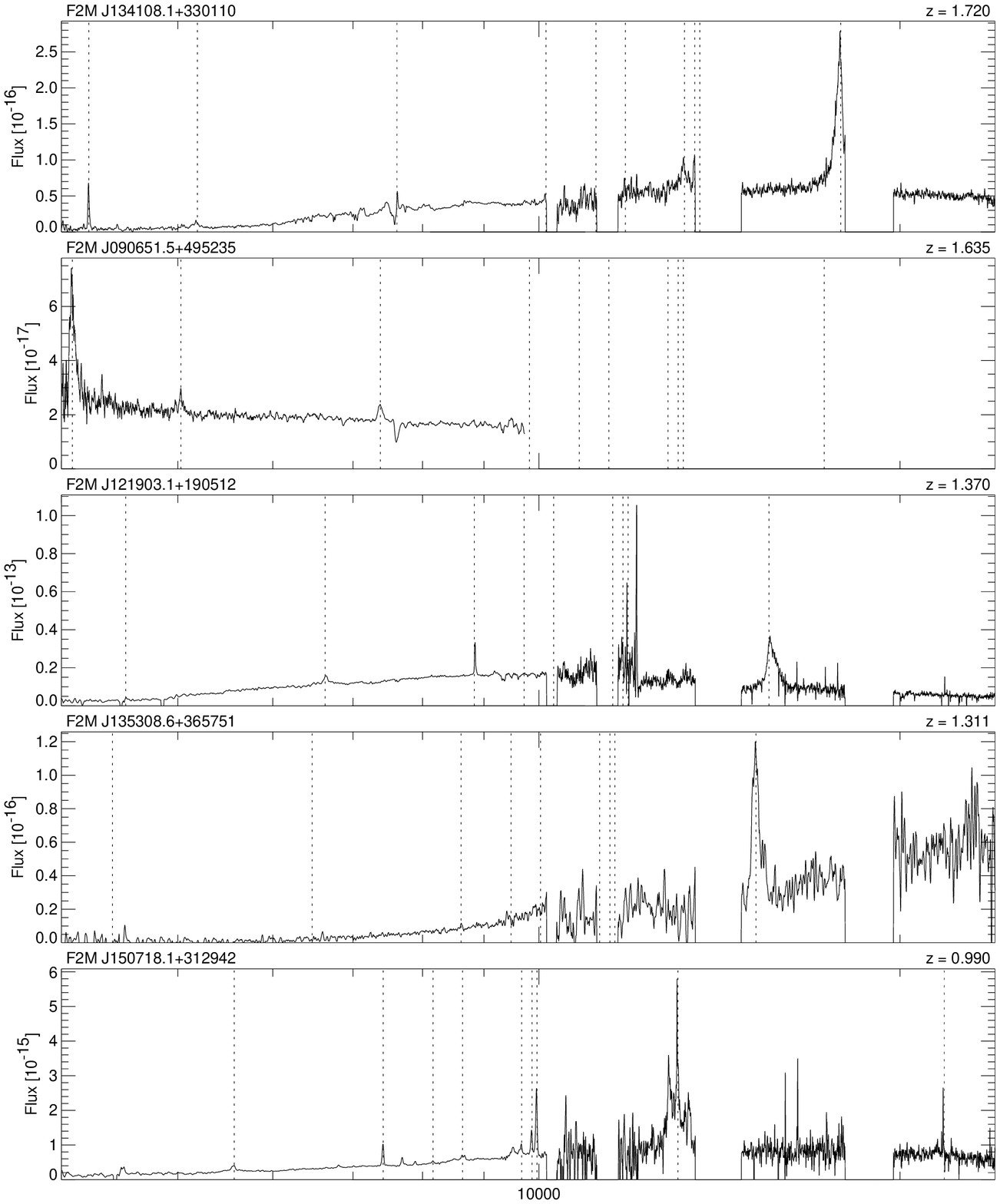}
\caption{{\it Continued.} Spectra of F2M quasars.}
\end{figure}

\begin{figure}
\figurenum{3c}
\plotone{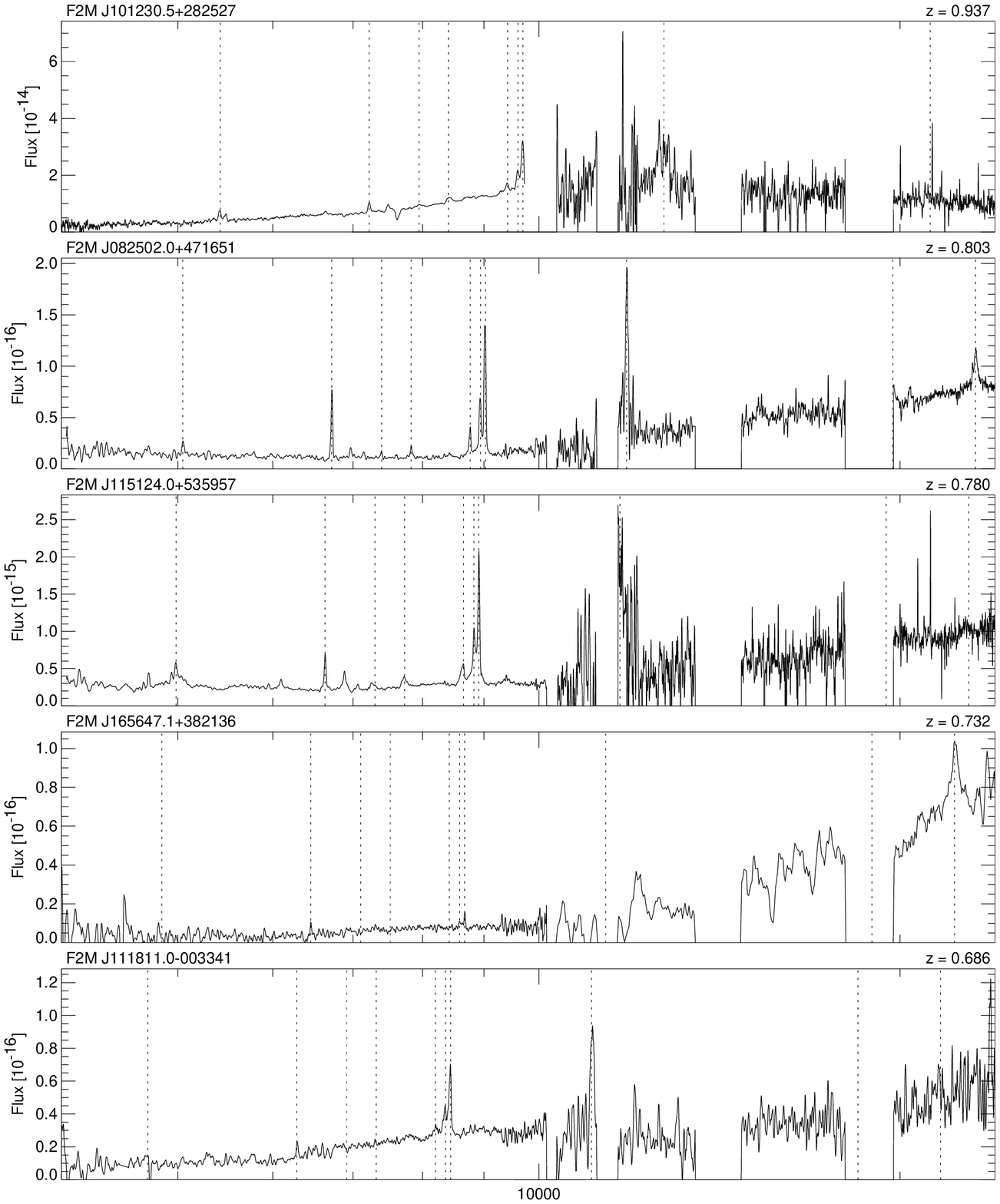}
\caption{{\it Continued.} Spectra of F2M quasars.}
\end{figure}

\begin{figure}
\figurenum{3d}
\plotone{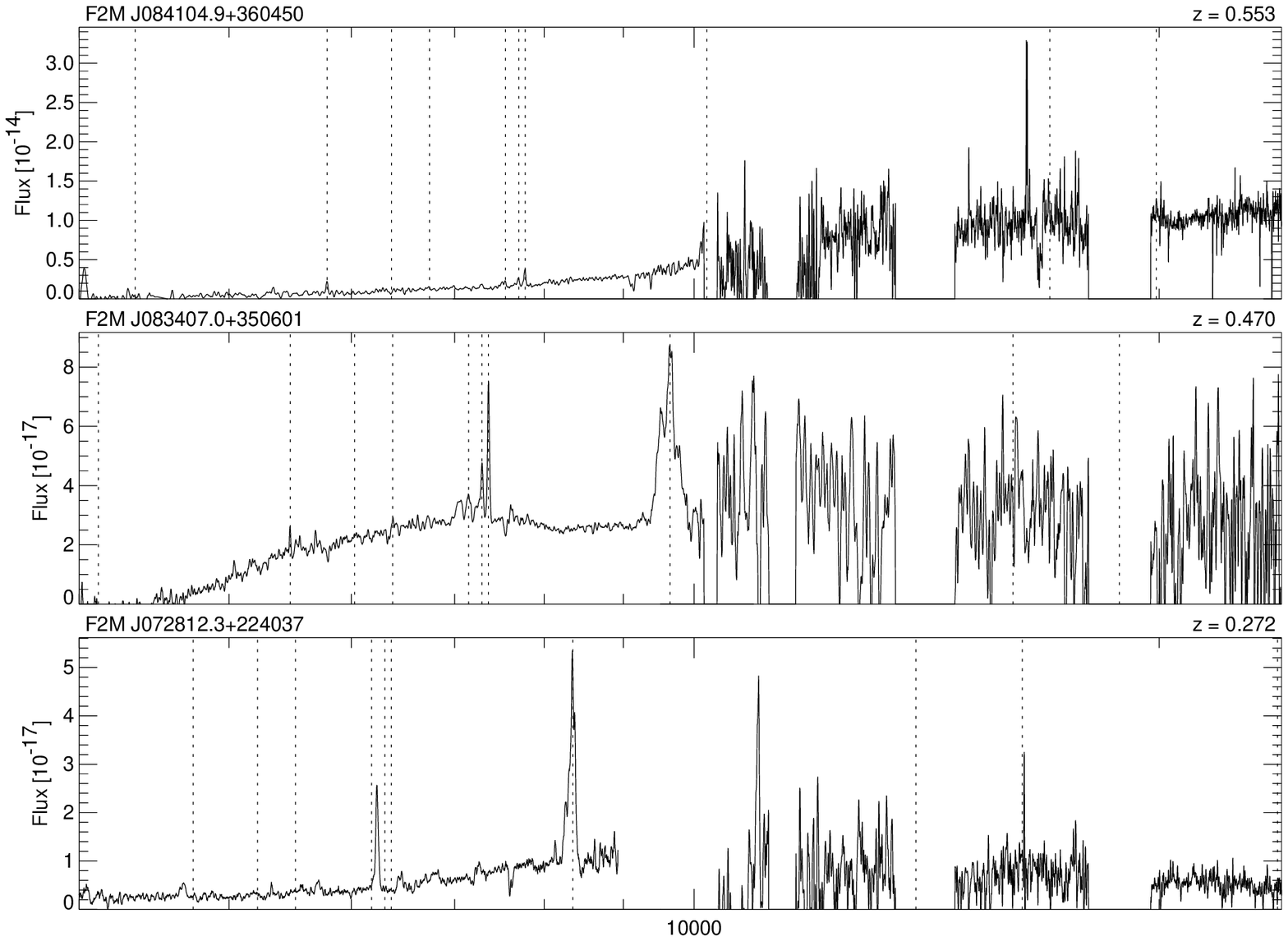}
\caption{{\it Continued.} Spectra of F2M quasars.}
\end{figure}

\begin{figure}
\figurenum{4}
\plotone{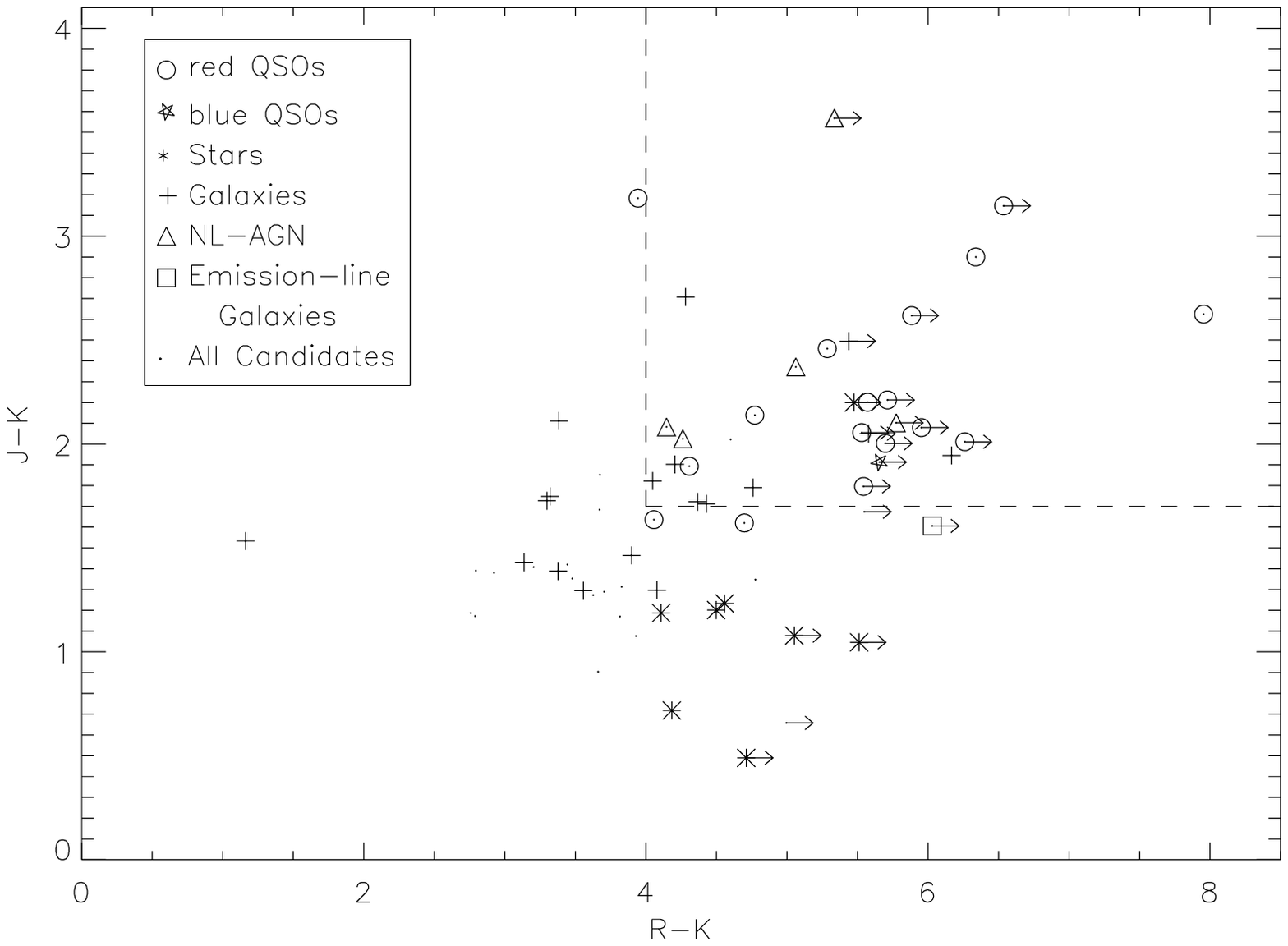}
\caption{Colors of our optically faint F2M objects plotted in $J-K$ vs. $R-K$ space.  Spectroscopic identifications are indicated.  The dashed lines define the region of this color-color space in which $\sim 50\%$ of the objects are red quasars.}
\end{figure}

\begin{figure}
\figurenum{5}
\plotone{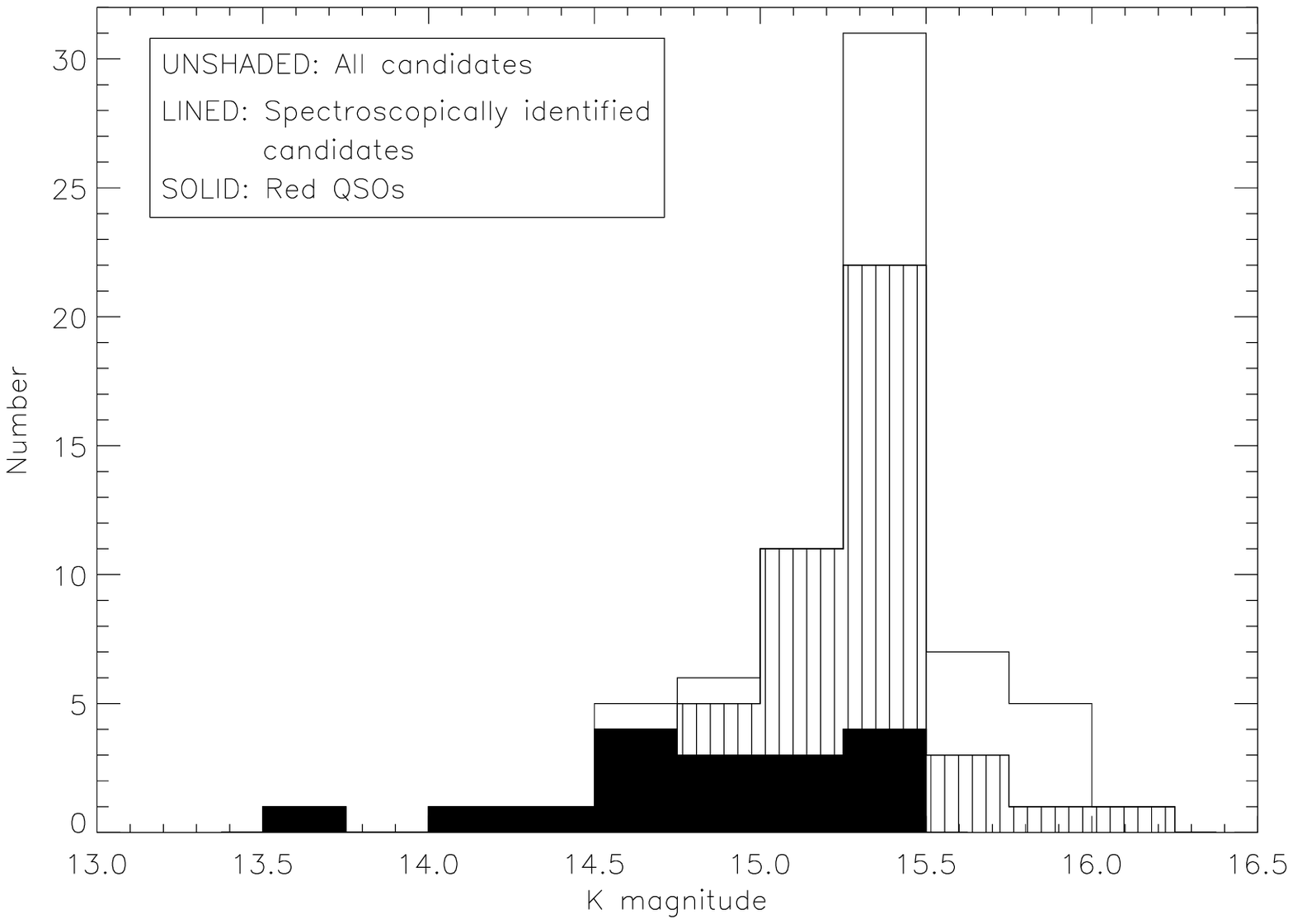}
\caption{Histogram of the $K$ magnitudes of the 69 candidates listed in Table 2 (unshaded).  All identified objects are overplotted in the shaded region, while the red quasars are overplotted with solid-colored bins.} 
\end{figure}

\begin{figure}
\figurenum{6}
\plotone{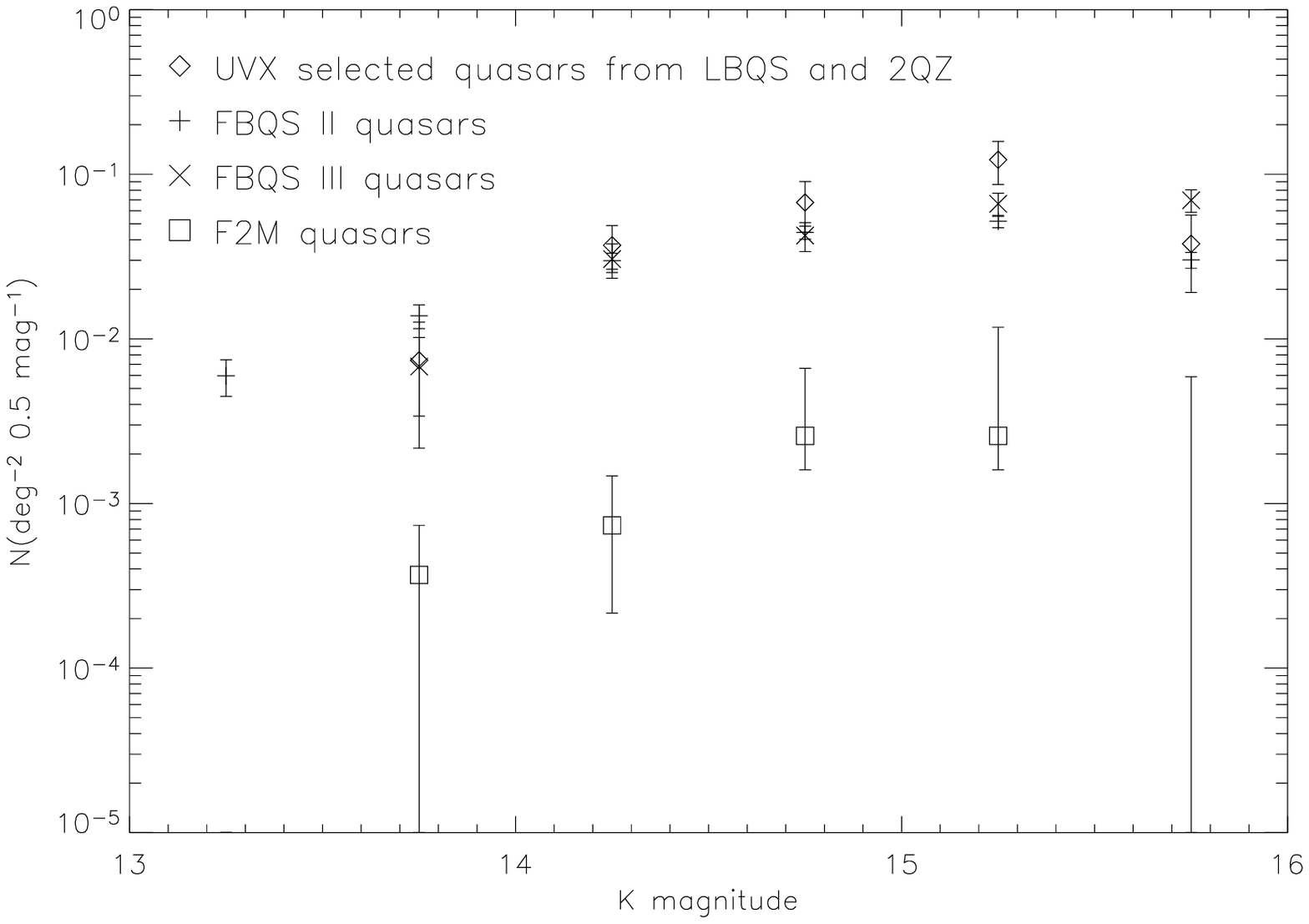}
\caption{Density of quasars on the sky from various surveys as a function of $K$ magnitude as listed in Table 3. The error bars for F2M red quasars include an upper bound which assumes all sources without spectroscopic data are quasars.  We add to these limits the usual $\sqrt{N}$ error bars.  All objects are detected in FIRST ($S_{20\mathrm{cm}} \geq 1$ mJy) and 2MASS ($K_s \leq 16$).} 
\end{figure}

\begin{figure}
\figurenum{7}
\plotone{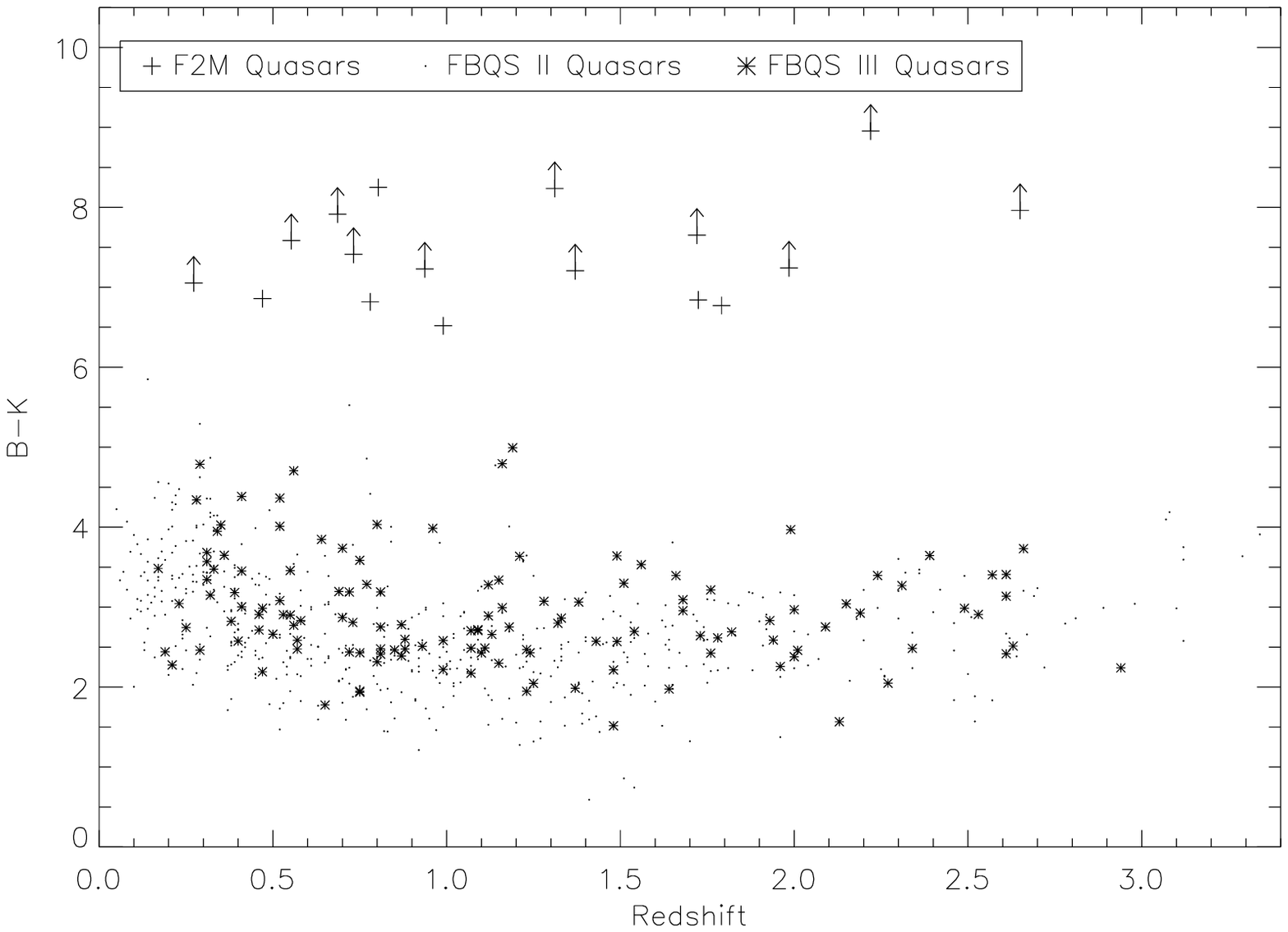}
\caption{Redshift distribution of F2M red quasars ($+$) compared with FBQS II ($\cdot$) and FBQS III ($\ast$).  Arrows indicate lower limits for objects not detected in the GSC 2.2 catalog.}
\end{figure}

\begin{figure}
\figurenum{8}
\plotone{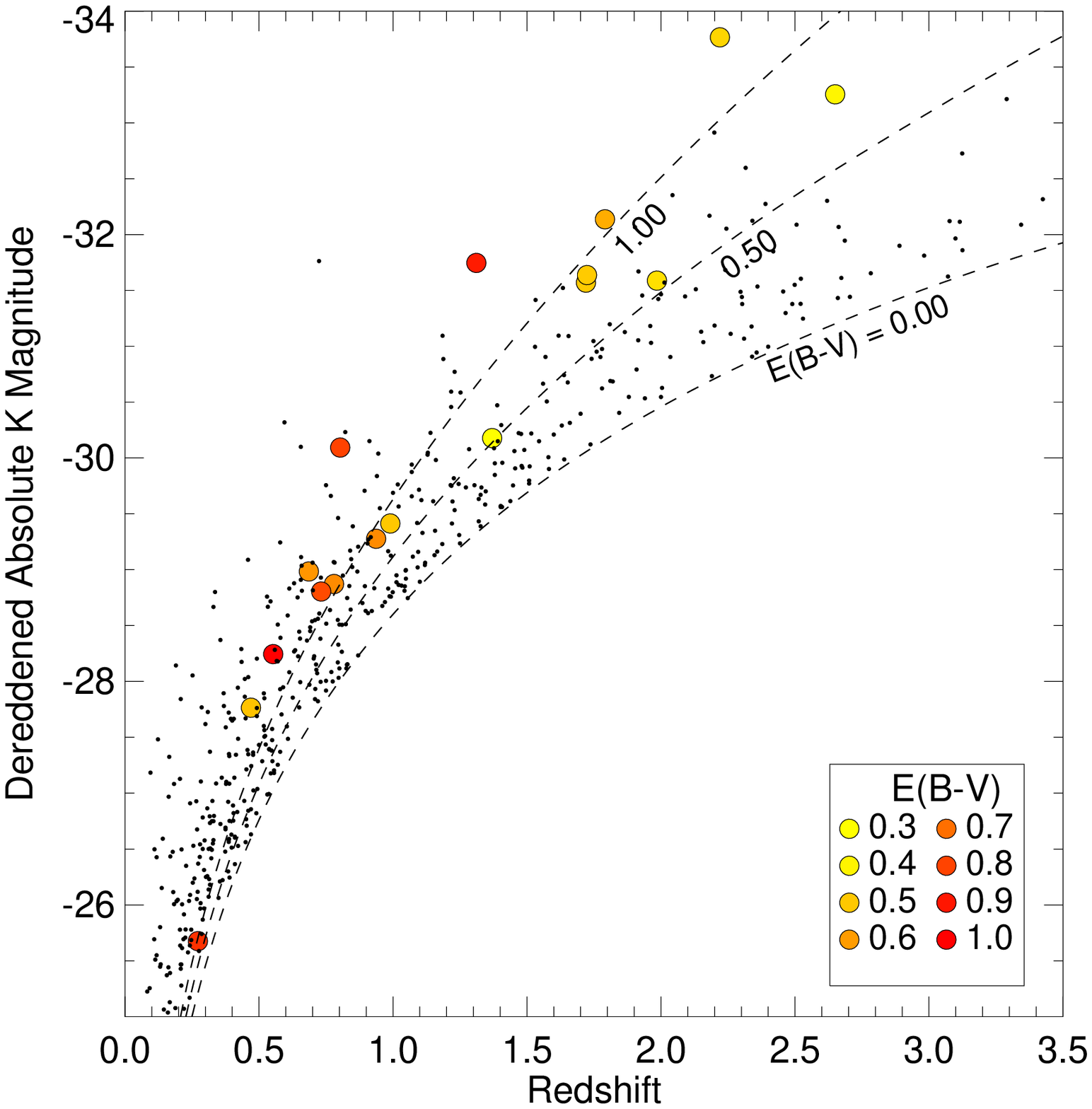}
\caption{Dereddened $K$-band absolute magnitude as a function of redshift in a magnitude-limited $K$-band survey.  The small dark points represent FBQS II and III quasars.  The shaded circles are the seventeen F2M red quasars.  Their colors correspond to the amount of extinction, ranging from $E(B-V)$ of $0.3-1.0$ where dark circles (redder colors in electronic edition) are more heavily reddened than lightly shaded (yellow colors in electronic edition) circles.  The dotted lines indicate the survey's $K < 15.5$ detection limit as a function of the amount of extinction, $E(B-V)$.  Note that the lwo highest luminosity, highest redshift objects are gravitationally lensed.} 
\end{figure}

\begin{figure}
\figurenum{9}
\plotone{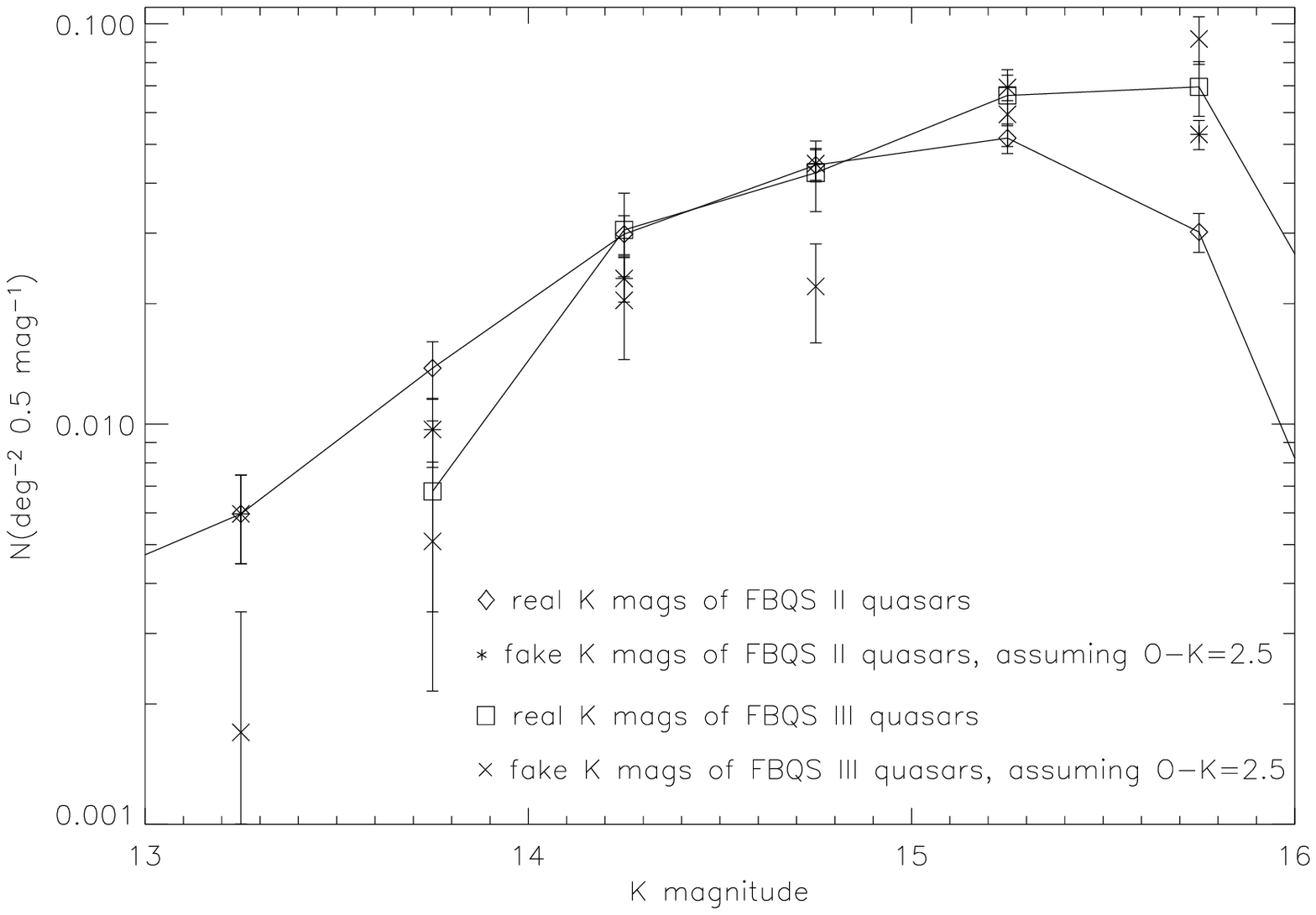}
\caption{This figure demonstrates that optically selected quasars have $B-K \simeq 2.5$.  The real points are 2MASS $K$ magnitudes matched to FBQS II ($\diamond$) and FBQS III ($\square$).  We compare these to ``fake'' points calculated from the APM $O$ magnitudes listed in the FBQS II and III catalogs, $K_{fake} = O - 2.5$ ($\ast$ and $\times$ symbols, respectively).  The real data are connected with solid lines to ease in distinguishing any differences between the real and fake points.}
\end{figure}

\begin{figure}
\figurenum{10}
\plotone{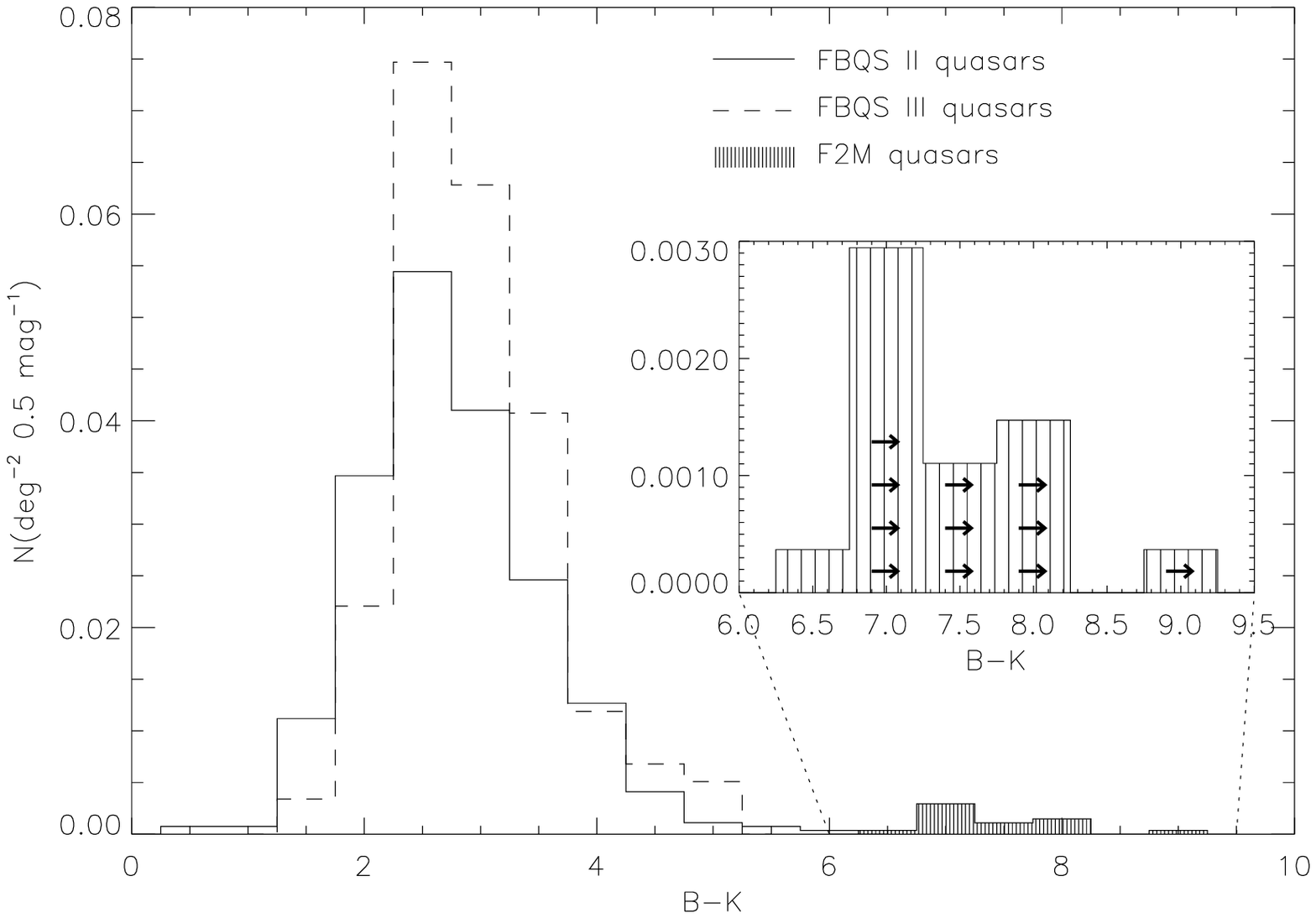}
\caption{$B-K$ color distribution for the FBQS II (solid line) and III (dashed line) quasars with 2MASS detections compared to the F2M sample (shaded).  The number counts have been scaled to each survey's respective coverage area to be comparable.  The inset plot magnifies the distribution of the F2M quasars.  Arrows indicate lower limits due to undetected $B$ magnitudes in the GSC2.2 catalog.}
\end{figure}

\end{document}